\pdfoutput=1
\documentclass[12pt,a4paper]{article}%
\usepackage{adjustbox}
\usepackage{comment}
\usepackage{multicol}
\usepackage{multirow}
\usepackage{lmodern}
\usepackage[T1]{fontenc}
\usepackage[utf8]{inputenc}
\usepackage{parskip}
\usepackage{amssymb}
\usepackage{amsfonts}
\usepackage{xfrac}
\usepackage{amsmath}
\usepackage{mathtools}
\usepackage{enumitem}
\usepackage[a4paper,margin=1in]{geometry}
\usepackage[figuresright]{rotating}
\usepackage[UKenglish]{isodate}
\usepackage{cases}
\DeclarePairedDelimiter\floor{\lfloor}{\rfloor}
\usepackage[authoryear]{natbib}
\usepackage[font=footnotesize,labelfont=sf]{caption}
\usepackage{setspace, tabularx}
\usepackage{booktabs}
\usepackage[flushleft]{threeparttable}
\usepackage[table,dvipsnames]{xcolor}
\usepackage{accents}
\newcommand\ubar[1]{%
  \underaccent{\bar}{#1}}
\usepackage{array}
\newcolumntype{L}{>{$}l<{$}}
\usepackage{abstract}

\usepackage[colorlinks=true,
            urlcolor=blue,
            citecolor=blue,
            linkcolor=blue,
            bookmarks,
            bookmarksopen=true,
            bookmarksnumbered=true,
            pdfstartview={FitH}
]{hyperref}
\newcommand{\Ex}{\textnormal{\textsf{E}}}

\newcommand{\cov}{\textnormal{\textsf{cov}}}
\newcommand{\corr}{\textnormal{\textsf{corr}}}

\newcommand{\eps}{\varepsilon}
\DeclareMathOperator*{\argmin}{\textnormal{\textsf{arg\,min}}}
\DeclareMathOperator*{\argmax}{\textnormal{\textsf{arg\,max}}}
\DeclareMathOperator*{\ssup}{\textnormal{\textsf{sup}}}
\DeclareMathOperator*{\mmax}{\textnormal{\textsf{max}}}
\DeclareMathOperator*{\plim}{\textnormal{\textsf{plim}}}

\newtheorem{assumption}{Assumption}
\newtheorem{lemma}{Lemma}
\newtheorem{proposition}{Proposition}
\newtheorem{corollary}{Corollary}
\usepackage[noblocks]{authblk}
\usepackage{graphicx}
\usepackage{tikz}
\usepackage{pgfplots}
    \usetikzlibrary{
        pgfplots.dateplot,
    }
\usepackage{subcaption}
\newcommand{\bk}{\color{black}}

 \usepackage{amssymb}
\usepackage{pifont}
\newcommand{\cmark}{\ding{51}}%
\newcommand{\xmark}{\ding{55}}%
\usepackage[table]{xcolor}
\definecolor{cream}{rgb}{1.0, 0.99, 0.82}
\makeatletter
\newcommand{\ccell}[3][]{%
  \kern-\fboxsep
  \if\relax\detokenize{#1}\relax
    \expandafter\@firstoftwo
  \else
    \expandafter\@secondoftwo
  \fi
  {\colorbox{#2}}%
  {\colorbox[#1]{#2}}%
  {#3}\kern-\fboxsep
}
\makeatother
\AtBeginDocument{\setlength{\cmidrulekern}{0.3em}}
\begin{document}


\title{Quantile Granger Causality in the Presence of Instability}
\author[1]{Alexander Mayer\thanks{Corresponding author. \emph{E-mail address}: \href{mailto:alexandersimon.mayer@unive.it}{alexandersimon.mayer@unive.it}. We would like to thank Lukas Hoesch for helpful comments and highly appreciate the suggestions and comments made by participants at the Econometrics Workshop 2024 at the Goethe University Frankfurt, the research seminar at Ca' Foscari University, Venice, and the IAAE 2024, Thessaloniki. We would like to thank the associate editor and two referees for valuable comments and suggestions that helped to improve the paper.}}
\author[2]{Dominik Wied}
\author[3]{Victor Troster}
\affil[1]{\small Università Ca' Foscari}
\affil[2]{\small University of Cologne}
\affil[3]{\small Universitat de les Illes Balears}
\date{\today}
\maketitle 
\thispagestyle{empty}

\begin{abstract}
\vspace{0.2in}
We propose a new framework for assessing Granger causality in quantiles in unstable environments, for a fixed quantile or over a continuum of quantile levels. Our proposed test statistics are consistent against fixed alternatives, they have nontrivial power against local alternatives, and they are pivotal in certain important special cases. In addition, we show the validity of a bootstrap procedure when asymptotic distributions depend on nuisance parameters. Monte Carlo simulations reveal that the proposed test statistics have correct empirical size and high power, even in absence of structural breaks. Moreover, a procedure providing additional insight into the timing of Granger causal regimes based on our new tests is proposed. Finally, an empirical application in energy economics highlights the applicability of our method as the new tests provide stronger evidence of Granger causality.\\

\end{abstract}

\newgeometry{}
\setcounter{page}{1}
\section{Introduction}

The definition of \cite{granger1969investigating} causality is a fundamental concept in time series econometrics. Accordingly, let $z_i$ denote a series contained within an information set that gathers all relevant information available up to time $i$, then $z_i$ is said to Granger-cause $y_{i}$ if $z_i$ provides information relevant to predicting $y_{i}$.  Although Granger causality is uncovered by the conditional distributions of $y_{i}$, applied research focuses often on Granger causality in mean because it entails easily testable implications. However, by solely testing the significance of $z_i$ in a conditional mean regression of $y_{i}$ on $z_i$, one runs the risk of neglecting possible tail relationships or nonlinearities.

It is for this reason that more recent research is also concerned with Granger causality in quantiles, which allows for an equivalent characterization of Granger causality in distribution. 
This implies that the conditional quantile function of $y_i$ depends on $z_{i}$ for some quantiles of interest, given all the available information until time $i$.  For instance, \citet{Lee2012} found fragile evidence of Granger causality between augmenting monetary policies and national income at the conditional mean; nevertheless, the authors reported strong evidence of Granger causality at extreme quantiles of the distribution.

One way to elicit potential evidence for Granger causality is by means of quantile regressions. \cite{koemach:99} developed a parametric significance test of quantile regression coefficients, which is frequently employed in empirical work to test for Granger causality in quantile regressions (see e.g. \citealp{Chuang2009a} or \citealp{Yang2014}). \cite{troster:18} extended the method of \cite{koemach:99} by providing a semiparametric omnibus test for Granger causality in quantiles that allows for nonlinear specifications of the quantile regressions under the null hypothesis of no Granger causality. On the other hand, \cite{Jeong2012}, \cite{Taamouti2014} and \cite{candelon2016} derived nonparametric tests for Granger causality in quantiles. \citet{bouezmarni_testing_2024} proposed such a test for expectiles.
 
What all these papers have in common, is, however, that they implicitly assume the pattern of Granger (non)causality to be stable over time. In this paper, we therefore propose tests for Granger causality in quantiles that are robust against temporal instabilities. This is of importance because financial and macroeconomic data structures, where Granger causality is frequently tested, are subject to strong fluctuations and volatility \citep{clark2006predictive,rossi:2005, rossi_advances_2013,roswan:19, stock_evidence_1996, stock_forecasting_1999, stock_forecasting_2003, stock_chapter_2006,baum_2021}. \cite{rossi2006exchange}, for instance, provides evidence of failure of traditional Granger causality tests to detect Granger causality from certain macroeconomic fundamentals to exchange rate fluctuations due to parameter instabilities in the models over time. \cite{chen2010can}, on the other hand, do report evidence of Granger causality from exchange rates to commodity prices, when applying Granger causality tests that allow for structural breaks. \citet{caporin:2022} have made similar arguments.  
 
In addition, \cite{giacomini2010forecast}, \cite{rossi_advances_2013}, and \cite{rossi2021forecasting}, among others, show that instabilities in the parameters of the models can affect the performance of Granger causality tests in different ways; hence, these authors recommend incorporating structural breaks in Granger causality tests rather than testing for instability in the parameters. Following this idea, it is important to apply methods that are robust to structural breaks or instabilities for correctly performing a Granger causality analysis in macroeconomic or financial time series.

To address potential temporal instabilities, we resort to the work of \cite{rossi:2005}, who --by extending the earlier work by \cite{sowell:96}-- developed tests for nested model selection with underlying parameter instability. Although the methods of \cite{rossi:2005} and, in particular, \cite{roswan:19} can be used to test for Granger causality in mean between two time series, they fall short to capture Granger causality in the tails or other parts of the conditional distribution not captured by the mean. The same is true for the time-varying Granger causality in-mean tests like the one employed, for example, by \cite{chen2010can} or \cite{caporin:2022}.   

We thus extend the method of \cite{rossi:2005} and \cite{roswan:19} in a consolidated way for testing for Granger causality in quantiles under structural instabilities. We do so by drawing from results on structural break testing in quantile regressions by \cite{qu:2008} and \cite{oka:11};  see also \cite{hoga:2024} for a recent extension to predictive quantile and CoVaR regression. To our knowledge, no test for Granger causality in quantiles with structural instability has been developed so far in the literature. 

More specifically, the main idea is to consider under the null the intersection of two sub-hypotheses: {\it First}, we hypothesize that the effect of the potential Granger-causing variate is constant over time and over quantiles. {\it Second}, we assume that this effect is zero. 
Likewise, the alternative in a local neighbourhood around the null consists of  two parts: one that specifies local deviations from the null hypothesis of no Granger causality in quantiles and another that specifies local deviations from the null hypothesis of constant parameters (over one or more quantiles of the distribution).
Thus, we construct our test statistics in such a way that ensure non-trivial local power against the union of these two alternative hypotheses. For this reason, we do not require {\it a priori} knowledge of whether any of the two alternative sub-hypotheses holds (or whether both hold). Finally, we propose tests that neither involve trimming over time nor require the specification of tuning parameters. Here we distinguish two cases, on the one hand, we have a single quantile, on the other hand, we consider a continuum of quantiles (as a subset of the interval $[0,1]$).

This idea goes back to, {\it inter alia}, \cite{sowell:96} and \cite{rossi:2005} who have developed similar test procedures in a general  generalized method of moments (GMM) framework.
Our extension is non-trivial. Unlike the GMM framework considered by \cite{rossi:2005}, we demonstrate, for example, that the limiting distribution might not be pivotal in certain cases. It is pivotal in certain important special cases such as homoskedasticity or conditional mean independence between regressors. In cases where the limiting distribution depends on nuisance parameters, we propose alternatives, including a new bootstrap procedure whose validity is established. This extends the semiparametric bootstrap used in \cite{rw:13}, where the estimated quantile functions are applied to uniformly distributed random variables. Therefore, we provide (bootstrap) test statistics with correct asymptotic size, which are consistent against fixed alternatives and possess nontrivial power against local alternative hypotheses. This is also corroborated by our Monte Carlo simulations; the finite-sample evidence shows that the test has appealing size and power properties in finite samples. In all cases (structural breaks and no structural breaks under the alternative), our tests are more powerful than the existing sup Wald test. Given that our new test rejects, the question arises how to interpret the result. To address this question, we propose an additional two-step procedure to further characterize the data generating process. After applying a plain CUSUM test for detecting and dating a structural break, we apply our new test in the detected regimes. So, we get precise information about in which parts of the time period we have Granger causality.

We illustrate the applicability of our tests by performing an empirical application. We revisit an analysis about the causal relationships between crude oil and stock returns from \cite{dingetal:16}, who consider the interplay between stock returns and crude oil returns. In the application, we find several scenarios which demonstrate the higher efficiency of the new test. Moreover, using our sequential procedure, we are able to identify economically interpretable regimes of Granger causality.

The rest of the paper proceeds as follows. In Section 2, we propose our test statistics for jointly testing for Granger causality in quantiles and parameter instability. In Section 3, we derive the asymptotic distribution of our test statistics; we also propose and justify a bootstrap method for implementing our test statistics. In Section 4, we perform Monte Carlo simulations to validate the finite-sample performance of our test statistics. In Section 5, we present an empirical application of our proposed tests. Finally, we conclude the paper in Section 6.

Throughout the paper, we use the following notation: ${\cal B}_m(\lambda)$, $\lambda \in [0,1]$, is a vector of $m$ independent Brownian motions, and ${\cal BB}_m(\lambda) \coloneqq {\cal B}_m(\lambda)-\lambda {\cal B}_m(1)$ is a vector of $m$ independent Brownian bridges. For a positive definite matrix $A$, $A^{-1/2}$ is defined as the Cholesky factor of its inverse $A^{-1}$, so that $A^{-1} = (A^{-1/2})'A^{-1/2}$. Notation ``$\Rightarrow$'' and ``$\rightarrow_d$'' indicates weak convergence and convergence in distribution, respectively. Notation $\mathcal{T}$ represents 
a closed interval such that $\mathcal{T} \subset [0,1]$. For an $m \times 1$ vector $z$, we define $ \Vert z \Vert_\infty \coloneqq \mmax\limits_{1 \leq j \leq m} |z_j|$. 
 
\section{Granger Causality}

Suppose we suspect that the $p \times 1$ vector $z_i$ Granger causes the dependent variable $y_i$, and, at the same time, we have reasons to question temporal stability. In other words, we expect Granger causality, but we are unsure whether its pattern persists over time.

Within the framework of a linear quantile regression, these considerations amount to parametrise the $\tau$ quantile of $y_i$ via
\begin{equation}
Q_{y_i}\left(\tau \mid x_i\right) \coloneqq x_i' \beta_i(\tau), \quad x_i \coloneqq (z_i',w_{i}')', \quad \beta_i(\tau) \coloneqq (\gamma_{i}(\tau)',\alpha(\tau)')' \in \mathbb{R}^{m},
\end{equation}
where $w$ is a $k \times 1$ vector of additional controls so that $m = p + k$, $\gamma_{i}(\tau) \neq 0$ for some $\tau \in [0,1]$, and $i \in \{1,\dots,n\}$ under Granger causality. For simplicity, we assume that all the available information up to time $i$ can be represented by vector $x_i$. For example, $x_i$ might be equal to $(y_{i-1},a_i,a_{i-1})$ for some univariate time series $a_i$. Two important models, which lead to such a structure, are the location-scale model, $y_i = x_i'\delta + (x_i'\rho) \eps_i$,
and the random coefficient model, $y_i = x_i'\beta(U_i)$, with a standard uniformly distributed random variable $U_i$ that includes, among others, quantile autoregressive distributed lag models (see e.g. \citealp{koenker:05} and \citealp{galvao:13}).

We formulate the following (joint) null hypothesis $H_0 \coloneqq H_{0,1}\cap H_{0,2}$:
\begin{eqnarray*}
H_{0,1} &\coloneqq& \left\{\gamma_{i}(\tau) = \gamma_{0}(\tau),\;\forall \mbox{ }i \in \{1,\dots,n\}, \tau \in \cal T \right\} \\
H_{0,2} &\coloneqq& \{\gamma_{0}(\tau) = 0_p, \forall \mbox{ } \tau \in\cal T\}
\end{eqnarray*}
against the alternative hypothesis $H_1 \coloneqq \{\neg H_{0,1}\} \cup \{\neg H_{0,2}\}$. More specifically, in a local neighbourhood around $H_0$ in the direction of $H_{1}$, the following sequence of local alternatives is investigated
\begin{align}\label{localt}
\gamma_{i,n}(\tau) \coloneqq \gamma_{0,n}(\tau) + \frac{\delta(\tau)}{\sqrt{n}}g\left(\frac{i}{n}\right),\quad \gamma_{0,n}(\tau) \coloneqq \frac{\Delta(\tau)}{\sqrt{n}},
\end{align}
where $\tau \mapsto \Delta(\tau)$ and $\tau \mapsto \delta(\tau)$ are deterministic continuous vector-valued and scalar-valued functions, respectively, and $v\mapsto g(v)$ is a deterministic vector-valued Riemann–Stieltjes integrable function. 

Our setup is essentially similar to the approach of \cite{rossi:2005} that extends the earlier procedure of \cite{sowell:96}, in which the null hypothesis consists of two different restrictions. On the one hand, the parameter $\gamma_i(\tau)$ is constant over $i$ and $\tau$; on the other hand, this constant is equal to $0$. Analogously, the alternative hypothesis also consists of two parts. The alternative $H_{1,1} \coloneqq \neg H_{0,1}$ specifies local deviations from the null hypothesis of constant parameters, whereas the alternative $H_{1,2} \coloneqq \neg H_{0,2}$ specifies local deviations from the null hypothesis of no Granger causality in quantiles. Our tests are constructed in such a way that they have power against the union of these alternatives. For this purpose,  it is not required to know {\it a priori} which of the two alternatives (or both) holds. 

\section{Test Statistics}
Consider the following sequential process based on the subgradient of the unconstrained quantile regression 
\begin{align}\label{eq:S_process}
S_{n}(\lambda,\tau,t) \coloneqq n^{-1/2}\sum_{i=1}^{\floor{\lambda n}}x_i\psi_\tau(y_i - x_i't), \quad t \in \mathbb{R}^{m},
\end{align}
where $\lambda \in [0,1]$ indexes the time fraction, and $\psi_\tau(u) \coloneqq 1\{u \leq 0\}-\tau$. Moreover, introduce the (constrained) estimator $\tilde\beta_n(\tau) \coloneqq \left(0, \alpha_{n}(\tau)'\right)'$, where
    \begin{align}\label{eq:beta_tilde}
    \alpha_{n}(\tau) \coloneqq \argmin\limits_{\alpha \in \mathbb{R}^k} \sum_{i=1}^n\rho_{\tau}(y_i-w_{i}'\alpha), \quad \rho_\tau(u) \coloneqq u(1\{u\leq0\}-\tau).
    \end{align}
Our tests \bk are \bk based on the following process
\begin{equation}\label{eq:Hn}
H_n(\lambda,\tau,t) \coloneqq (X_n'X_n/n)^{-1/2}S_n(\lambda,\tau,t),\quad \lambda,\tau \in [0,1],\;t \in \mathbb{R}^m,
\end{equation}
where $X_n \coloneqq (x_1',\dots,x_n')'$ is $n \times m$. As pointed out by \cite{qu:2008}, the process in\ \eqref{eq:Hn} is asymptotically pivotal when evaluated at the true parameter vector; see also \cite{parzen:94} for a similar argument. This allows us to construct tests that do not require trimming over time.

The main idea behind our test statistic is to combine two detectors that are respectively designed to find deviations from $H_{0,1}$ and $H_{0,2}$. More specifically, a CUSUM-type statistic 
$${\sf LM_1}(\lambda,\tau) \coloneqq \left\Vert R' \Delta H_n(\lambda,\tau,\tilde\beta_n(\tau)) \right\Vert_\infty, \quad R \coloneqq \begin{bmatrix} I_p \\ 0_{k\times p} \end{bmatrix},$$
with
$$
\Delta H_n(\lambda,\tau,\tilde\beta_n(\tau)) \coloneqq H_n(\lambda,\tau,\tilde\beta_n(\tau))-\lambda  H_n(1,\tau,\tilde\beta_n(\tau))
$$
is used to test $H_{0,1}$, while the restriction of $H_{0,2}$ is verified using the LM (Lagrange Multiplier) statistic
$$ {\sf LM_2}(\tau) \coloneqq \left\Vert R'H_n(1,\tau,\tilde\beta_n(\tau)) \right\Vert_\infty.$$
Since ${\sf LM}_1$ has no power against constant deviations from the null, \bk and \bk ${\sf LM}_2$ lacks power if Granger causality is unstable, our tests will be of the form 
\begin{equation}\label{psi}
\varphi\big({\sf LM_1}+{\sf LM_2}\big),
\end{equation}
for some weighting function $\varphi: [0,1] \times [0,1] \mapsto \mathbb{R}$ specified below.

In what follows, we distinguish between situations where our interest lies in detecting deviations from the null ($i$) at a given quantile or ($ii$) across various quantiles. \bk To \bk derive the properties of theses tests, we impose the following assumptions that are similar to those in \cite{qu:2008} and \cite{oka:11}.

\renewcommand{\theassumption}{A}
\begin{assumption}\label{ass:A} Let $u_{i,n}(\tau) \coloneqq y_i-\beta_{i,n}(\tau)'x_i$. Then $1\{u_i(\tau)\leq 0\}-\tau$ is a martingale difference array with respect to $\mathcal F_{i-1}\coloneqq \sigma(\{y_{j-1},x_j: j \leq i\})$ for any $\tau \in [0,1]$.
\end{assumption}

Let $f_i(\cdot)$, $F_i(\cdot)$ and $F^{-1}_i(\cdot)$ denote the conditional density, conditional distribution, and conditional quantile function, respectively, of $y_i$ given $w_i$.

\renewcommand{\theassumption}{B}
\begin{assumption}\label{ass:B} 
\textcolor[rgb]{1,1,1}{.}
\begin{enumerate}[label=\textnormal{\textbf{B.\arabic*}},ref=B.\arabic*]
    \item The distribution functions $F_i(\cdot)$ are absolutely continuous, with continuous densities $f_i(\cdot)$ satisfying $0 < \ubar{u} \leq f_i(F^{-1}_i(\tau)) \leq \bar{u} < \infty$ for all $i$.
    \item For any $\epsilon > 0$, there exists a $\sigma(\epsilon) > 0$ such that
$|f_i(F^{-1}_i(\tau)+s)-f_i(F^{-1}_i(\tau))| \leq \epsilon$ for all $|s| < \sigma(\epsilon)$ and all $1 \leq i \leq n$.
\end{enumerate}
\end{assumption}

\renewcommand{\theassumption}{C}
\begin{assumption}\label{ass:C} 
The regressors are assumed to satisfy: 
\textcolor[rgb]{1,1,1}{.}
\begin{enumerate}[label=\textnormal{\textbf{C.\arabic*}},ref=B.\arabic*]
    \item The vector $w$ contains a constant.
\item  $\plim_{n \rightarrow \infty} \frac{1}{n}\sum_{i=1}^{\floor{\lambda n}}f_i(F^{-1}_i(\tau))x_ix_i' = \lambda {\sf H}(\tau)$ uniformly in $\lambda \in [0,1]$, where ${\sf H}(\tau)$ is a $m \times m$ non-random positive definite matrix.
\item There exists $a > 0$ and $A<\infty$ such that $\Ex[\left\Vert x_i\right\Vert^{4+a}] \leq A$.  
\item There exists $b > 0$ and $B<\infty$ such that for any $n$:
$$\frac{1}{n}\sum_{i=1}^n\Ex[\left\Vert x_i\right\Vert^{3(1+b)}] \vee \Ex[\frac{1}{n}\sum_{i=1}^n\left\Vert x_i\right\Vert^{3}]^{1+b} \leq B.$$
\item $\plim_{n \rightarrow \infty} \frac{1}{n} \sum_{i=1}^{\floor{\lambda n}}x_ix_i' = \lambda {\sf J}$ uniformly in $\lambda \in [0,1]$, where ${\sf J}$ is a $m \times m$ non-random positive definite matrix.
\end{enumerate}
\end{assumption}

These assumptions are standard in the context of tests for structural breaks in quantile models and of tests for Granger causality in quantiles. Assumption \ref{ass:A} restricts the dependence over time. Serial independence is not required, instead we have a martingale difference assumption on the innovations. Assumption \ref{ass:B} introduces positivity and smoothness assumptions on the conditional density of $y_i$ given $x_i$. Assumption C imposes restrictions on the regressors $x_i$, in particular on the existence of moments. This assumption rules out trends in the regressors, but it allows for heteroscedasticity. It might be interesting to relax this assumption in future work to allow also for frequency-dependent regressors as, for example, in \cite{li:08} in order to test for Granger causality at different frequencies akin to \cite{breitungcandelon:2006}.

\subsection{Granger Causality at a Given Quantile}

Let $\beta_0(\tau) \coloneqq (0_p',\alpha_{0}(\tau)')'$ be the true coefficient under the null and define ${\sf C}(\tau) \coloneqq {\sf J}^{-1/2}{\sf H}(\tau).$ Note that ${\sf C}(\tau)$ is a square root of the inverse of the variance-covariance matrix of the limiting distribution of the estimator that solves the unrestricted quantile regression problem (see, e.g. \citealp{koenker:05}). Moreover, let ${\sf H}_{\alpha}(\tau)$ and ${\sf J}_{\alpha}$, denote the lower-right $k \times k$ block of {\sf H}$(\tau)$ and {\sf J}, respectively, and partition 
$\beta_0(\tau) = (\gamma_{0}(\tau)',\alpha_{0}(\tau)')'$, where, under the null, $\gamma_0(\tau) = 0_p$.
 
The limiting distribution of the restricted quantile estimator, and the two detectors can now be summarized as follows.

\begin{proposition}\label{corHtilde} 
Assume that Assumptions \ref{ass:A}, \ref{ass:B}, and \ref{ass:C} hold. For a given $\tau \in {\cal T}$, 
\[
{\sf C}(\tau)\sqrt{n}(\tilde\beta_n(\tau)-\beta_0(\tau)) = -{\sf P}(\tau){\sf J}^{-1/2}S_{n}(1,\tau,\beta_{0}(\tau))  + o_p(1),
\]
where 
\[
P(\tau) \coloneqq {\sf C}(\tau) \bar R(\bar R'{\sf C}(\tau) \bar R)^{-1}\bar R', \quad
\bar R \coloneqq \begin{bmatrix} 0_{p\times k} \\ I_{k} \end{bmatrix}.
\]
Moreover, it holds that uniformly in $\lambda \in [0,1]$
\[
 h(\tau){\sf J}^{-1/2}S_n(\lambda,\tau,\beta_0(\tau)) \Rightarrow {\cal B}_m(\lambda)+  h(\tau){\sf C}(\tau) R\left(\lambda  \Delta(\tau)+\delta(\tau) \int_0^\lambda g(v){\sf d}v\right),
 \]
 $h^2(\tau) \coloneqq 1/(\tau(1-\tau))$, so that
\begin{align*}
  h(\tau)& R'\Delta H_n(\lambda,\tau,\tilde\beta_n(\tau))\\  &\Rightarrow  {\cal BB}_p(\lambda) + h(\tau)\delta(\tau) R'{\sf C}(\tau)R\left((1-\lambda)\int_0^\lambda g(v){\sf d}v  -\lambda \int_\lambda^1 g(v){\sf d}v\right) \\
  \,& \qquad \eqqcolon {\cal Z}^{(1)}(\lambda,\tau), 
\end{align*}
and 
\begin{align*}
h(\tau)&R'H_n(1,\tau,\tilde\beta_n(\tau))\\ 
\,&\Rightarrow 
    \tilde {\cal B}_p(1,\tau) +
h(\tau)
    R'T(\tau){\sf C}(\tau)R\left(\Delta(\tau)+\delta(\tau)\int_0^1g(v){\sf d}v\right)  \eqqcolon {\cal Z}^{(2)}(\tau),
\end{align*}
where $\tilde{\cal B}_m(\lambda,\tau) \coloneqq T(\tau){\cal B}_m(\lambda)$, with $T(\tau)$ denoting the inverse of the $m \times m$ matrix of eigenvectors of $I-P(\tau)$.
\end{proposition}

Interestingly, and contrary to the corresponding GMM result in \cite{rossi:2005}, the limiting distribution of the LM statistic $H_n(1,\tau,\tilde\beta_n(\tau))$ is not pivotal because the projection matrix $P(\tau)$ is oblique (i.e. idempotent of rank $k$ but not symmetric). An important exception is given if the following additional condition is satisfied: 
 
\renewcommand{\theassumption}{D}
\begin{assumption}\label{ass:D}  The  $p \times k$ matrix $Q(\tau) \coloneqq R'{\sf C}(\tau)\bar R {\sf H}^{-1}_{\alpha}(\tau){\sf J}^{1/2}_{\alpha}$ is zero.
\end{assumption}
 
Assumption\ \ref{ass:D} ensures that the oblique projection matrix $P(\tau)$ defined in Proposition\ \ref{corHtilde} is equal to the orthogonal projection $\bar R(\bar R'\bar R)^{-1}\bar R' = \bar R\bar R'$, which follows from observing that $P(\tau)$ decomposes into an orthogonal projection perturbed by a nilpotent matrix
\begin{equation}\label{project}
P(\tau) = {\sf C}(\tau) \bar R(\bar R'{\sf C}(\tau) \bar R)^{-1}\bar R' = \bar R\bar R'  + \begin{bmatrix}
    0_{p\times p} & Q(\tau) \\ 0_{k\times p} & 0_{k\times k}
\end{bmatrix}.
\end{equation}
A sufficient condition for Assumption\ \ref{ass:D} is ${\sf H}c(\tau) = {\sf J}(\tau)$ for some scalar $c(\tau) \in (0,\infty)$, which holds under homoscedasticity. Alternatively, Assumption\ \ref{ass:D} is satisfied if $w$ is just a constant or, more generally, under conditional mean independence of $z$ with respect to $w$ (i.e. $\Ex[z \mid w] = \Ex[z]$) as both of theses conditions ensure under the null $Q(\tau) = 0_{p \times k}$.

\begin{corollary}\label{corH0}   
Under the null hypothesis and Assumptions \ref{ass:A}--\ref{ass:D}, we get for a given $\tau \in {\cal T}$ and uniformly in $\lambda \in [0,1]$ 
\[
h(\tau)H_n(\tau,\lambda,\tilde\beta_n(\tau)) \Rightarrow \begin{bmatrix} {\mathcal B}_p(\lambda) \\ {\cal BB}_k(\lambda) \end{bmatrix},
\]
where ${\cal BB}_k(\cdot)$ and ${\cal B}_p(\cdot)$ are independent.
\end{corollary}

The $p$-dimensional Brownian motion ${\cal B}_p$ and the $k$-dimensional Brownian bridges ${\cal BB}_k$ arise due to the restricted and unrestricted components of the process tow-parameter process $H_n(\lambda,\tau,\beta(\tau))$, respectively. The limiting random variable under the null is independent of $\tau$. While this is true if Assumption\ \ref{ass:D} holds, violations from this assumption introduce dependence on nuisance parameters (cf. Proposition \ref{cor1}).

Based on the previous result, we will now introduce our first test statistic, suited to test $H_0$ at a given quantile $\tau$
\begin{align*}\label{eq:GCtest1}
   {\sf LM}(\tau) \coloneqq h(\tau)\left(\ssup\limits_{\lambda \in [0,1]}{\sf LM}_{1}(\lambda,\tau) + {\sf LM}_{2}(\tau)\right).
\end{align*}

For fixed $\tau$, the test statistic essentially consists of the sum of two individual test statistics, which reflect the two parts of the alternative hypothesis. Both statistics are based on the standardized subgradient of the unconstrained quantile regression through the process $H_n(\lambda,\tau,t)$ from \eqref{eq:Hn}. The first part, ${\sf LM}_{1}(\lambda,\tau)$, is the CUSUM part that detects structural breaks in the parameter $\gamma_i(\tau)$. Typically, for CUSUM statistics, one considers the supremum over the potential breakpoints $\lambda \in [0,1]$. The second part, ${\sf LM}_{2}(\tau)$, is essentially the LM  statistic for the hypothesis $H_{0,2}$.  

Corollary \ref{corHtilde} states why it makes sense to consider the sum of the two individual test statistics: The first statistic does have local power against structural breaks, but it has no power if there is Granger causality with constant parameters. The second part has power if there is Granger causality with constant parameters, but it has no power if $\Delta(\tau) = 0$ and $\int_0^1g(v){\sf d}v = 0$. The interpretation of the latter would be that there are structural breaks that lie in opposite directions over time.

\begin{corollary}\label{cor1}
For a given $\tau \in {\cal T}$, we get under the assumptions of Proposition \ref{corHtilde}
\[
{\sf LM}(\tau) \rightarrow_d 
\ssup\limits_{\lambda \in [0,1]} \left\Vert{\cal Z}^{(1)}(\lambda,\tau)\right\Vert_\infty+\left\Vert{\cal Z}^{(2)}(\tau)\right\Vert_\infty,
\]
while under the null
\[
{\sf LM}(\tau) \rightarrow_d 
\ssup\limits_{\lambda \in [0,1]} \left\Vert{\cal BB}_p(\lambda)\right\Vert_\infty+\left\Vert{\tilde{\cal B}}_p(1,\tau)\right\Vert_\infty.
\]
where ${\cal BB}_p$ and $\tilde {\cal B}_p(\lambda,\tau)$ are independent. If Assumption \ref{ass:D} holds, then $\tilde {\cal B}_p(\lambda,\tau) = {\cal B}_p(\lambda)$.
\end{corollary}

Thus, unless Assumption \ref{ass:D} is satisfied, the limiting distribution is not pivotal due to the second element ${\sf LM}_{2}$ of our test statistic that induces dependence on the quantile level $\tau$ via $Q(\tau)$. More specifically, it can be shown that
\[
{\sf LM}_{2}(\tau) = \left\Vert R'(I-P(\tau)){\sf J}^{-1/2}S_n(1,\tau,\beta_0(\tau))\right\Vert_\infty + o_p(1),
\]
where the oblique projection $P(\tau)$ causes quantile dependece because it cannot be diagonalized; cf. Eq.\ \eqref{project}. The distribution of ${\sf LM}_2$ can be viewed as a maximum of $p$-scaled absolute standard normals, where--similar to the discussion in \cite{hansen:21}--the scaling differs in general from unity, thereby capturing deviations from Assumption \ref{ass:D} (e.g. from homoskedasticity to heteroskedasticity).

\subsection{Granger Causality at all Quantiles}

To avoid multiple testing issues when performing inference across various quantiles, we extend the test statistics from the previous section to allow uniform inference across both $\lambda$ {\it and} $\tau$. Following \cite{andpol:1994} and  \cite{hansen:1996b}, we consider the following test statistics: 
\begin{equation}\label{eq:GCtest1}
\begin{split}
{sup\sf{LM}} \coloneqq \,&\ssup\limits_{\tau \in \cal T}\left(\ssup\limits_{\lambda \in [0,1]}{\sf LM}_{1}(\lambda,\tau) + {\sf LM}_{2}(\tau)\right), \\
   \textnormal{\it{exp}\sf{LM}} \coloneqq \,&\int_{\cal T}{\sf exp}\left[\frac{1}{2}\left(\ssup\limits_{\lambda \in [0,1]}{\sf LM}_{1}(\lambda,\tau) + {\sf LM}_{2}(\tau)\right)\right] d\tau. 
   \end{split}
\end{equation}
While both weighting schemes (over $\tau$) direct power against relatively distantly located alternatives, $\textnormal{\it{exp}\sf{LM}}$ can be considered optimal (see also \citealp{rossi:2005}).

Similar to Proposition \ref{corHtilde}, we first derive the properties of the process Eq. \eqref{eq:Hn} that serves as the building block of our test statistics. To this end, define $\widetilde{\cal S}_m(\tau,\lambda) \coloneqq T(\tau){\cal S}_{m}(\tau,\lambda)$, $T(\tau)$ defined in Proposition \ref{corHtilde}, ${\cal S}_{m}(\lambda,\tau) \coloneqq ({\cal S}_{1m}(\lambda,\tau),\dots,{\cal S}_{mm}(\lambda,\tau))'$ is an $m \times 1$ vector of independent Gaussian processes with
\[
\cov[{\cal S}_{im}(\lambda_1,\tau_1),{\cal S}_{im}(\lambda_2,\tau_2)] = (\lambda_1 \wedge \lambda_2)(\tau_1 \wedge \tau_2 - \tau_1\tau_2)
\]
and ${\cal SS}_m(\lambda,\tau) \coloneqq {\cal S}_m(\lambda,\tau) - \lambda {\cal S}_m(1,\tau)$ so that
\[
\cov[{\cal SS}_{im}(\lambda_1,\tau_1),{\cal SS}_{im}(\lambda_2,\tau_2)] = (\lambda_1 \wedge \lambda_2-\lambda_1\lambda_2)(\tau_1 \wedge \tau_2 - \tau_1\tau_2).
\]


\begin{proposition}\label{prop2} Assume that Assumptions \ref{ass:A}, \ref{ass:B}, and \ref{ass:C} hold uniformly in $\tau$. Then, uniformly in $\tau \in \mathcal{T}$, we have
\[
{\sf C}(\tau)\sqrt{n}(\tilde\beta_n(\tau)-\beta_0(\tau)) = -{\sf P}(\tau){\sf J}^{-1/2}S_{n}(1,\tau,\beta_{0}(\tau))  + o_p(1).
\]
Moreover, it holds that uniformly in $(\tau,\lambda) \in \mathcal{T} \times [0,1]$
\[
{\sf J}^{-1/2}S_n(\lambda,\tau,\beta_0(\tau)) \Rightarrow {\cal S}_m(\lambda,\tau)+  {\sf C}(\tau) R\left(\lambda  \Delta(\tau)+\delta(\tau) \int_0^\lambda g(v){\sf d}v\right),
 \]
so that 
\begin{align*}
R'\Delta H_n(\lambda,\tau,\tilde\beta_n(\tau)) \Rightarrow  \,& {\cal SS}_p(\lambda,\tau) + \delta(\tau) R'{\sf C}(\tau)R\left((1-\lambda)\int_0^\lambda g(v){\sf d}v  -\lambda \int_\lambda^1 g(v){\sf d}v\right) \\
\,& \eqqcolon {\cal Y}^{(1)}(\lambda,\tau)
\end{align*}
and
\begin{align*}
R'H_n(1,\tau,\tilde\beta_n(\tau))  \Rightarrow 
    \widetilde{\cal S}_p(1,\tau) +
    R'T(\tau){\sf C}(\tau)R\left(\Delta(\tau)+\delta(\tau)\int_0^1g(v){\sf d}v\right)  \eqqcolon {\cal Y}^{(2)}(\tau).
\end{align*}
\end{proposition}

Again, the weak limit of $H_n(1,\tau,\tilde\beta(\tau))$ is affected by nuisance parameters unless Assumption \ref{ass:D} holds, in which case $\widetilde {\cal S}_p(\lambda,\tau) = {\cal S}_p(\lambda,\tau).$ Moreover, note that ${\cal S}(1,\tau) = {\cal B}(\tau)$, while the Gaussian process ${\cal SS}(\lambda,\tau)$ is also referred to as a Brownian pillow or a pinned Brownian sheet; see also \citet[Sec 4]{qu:2008}. From the above, the limiting distribution of the test statistics follows readily by the continuous mapping theorem:

\begin{corollary}\label{cor4}
Uniformly in $(\tau,\lambda) \in \mathcal{T} \times [0,1]$, we have under the conditions of Proposition \ref{prop2}
\begin{align*}
   {sup\sf{LM}} \rightarrow_d \,&\ssup\limits_{\tau \in \cal T}\left(\ssup\limits_{\lambda \in [0,1]} \left\Vert {\cal Y}^{(1)}(\lambda,\tau) \right\Vert_\infty + \left\Vert{\cal Y}^{(2)}(\lambda,\tau) \right\Vert_\infty\right),  \\
   \textnormal{\it{exp}\sf{LM}} \rightarrow_d \,&\int_{\cal T}{\sf exp}\left[\frac{1}{2}\left(\ssup\limits_{\lambda \in [0,1]}\left\Vert {\cal Y}^{(1)}(\lambda,\tau) \right\Vert_\infty + \left\Vert {\cal Y}^{(2)}(\lambda,\tau) \right\Vert_\infty\right)\right] d\tau, 
\end{align*}
while, under the null, ${\cal Y}^{(1)}(\lambda,\tau) = {\cal SS}_p(\lambda,\tau)$ and ${\cal Y}^{(2)}(\lambda,\tau) = \widetilde {\cal S}_p(1,\tau)$.  If Assumption \ref{ass:D} holds, then $\widetilde {\cal S}_p(\lambda,\tau) = {\cal S}_p(\lambda,\tau)$.
\end{corollary}

\subsection{Practical Implementation}

 If Assumption \ref{ass:D} is satisfied, then it is easy to simulate the limiting distributions because they are free of unknown nuisance parameters (cf. Corollaries \ref{cor1} an \ref{cor4}). More specifically, using numerical techniques we can arbitrarily well approximate the Brownian motion ${\cal B}(\cdot)$ and the Brownian bridge ${\cal BB}(\cdot)$ for the fixed-$\tau$ case or, if instead a continuum of quantiles is considered, the limiting processes ${\cal S}(\cdot)$ and  ${\cal SS}(\cdot)$  (see the discussion in \cite{and:93} and \cite{qu:2008} for details on the numerical computation). 
 
If Assumption \ref{ass:D} is violated, then  we could still simulate the limiting distributions. But since the weak limits depend in this case on the characteristics of the {\it dgp} (cf. Corollaries \ref{cor1} an \ref{cor4}), we need to tabulate critical values for each application separately. In principle, one could proceed as follows: Firstly, based on a consistent estimator ${\sf H}_n(\tau)$, say, of ${\sf H}(\tau)$, we estimate $Q(\tau)$ using $Q_n(\tau) \coloneqq {\sf J}_n^{-1/2}{\sf H}_n(\tau){\sf H}_{n,\alpha}(\tau){\sf J}^{1/2}_{n,\alpha}$, with ${\sf J}_n \coloneqq X'X/n$, and where ${\sf H}_{n,\alpha}(\tau)$ and ${\sf J}_{n,\alpha}$ denote, respectively, the sample analogues of {\sf H}$_\alpha(\tau)$ and {\sf J}$_\alpha$ from Assumption \ref{ass:D}. For instance, a consistent estimator can be obtained via
\[
{\sf H}_n(\tau) = \frac{1}{2nc_n} \sum_{i=1}^n 1\{\hat{u}_i(\tau)\leq c_n\} x_ix_i',
\]
where $c_n \rightarrow 0$, $\sqrt{n}c_n \rightarrow \infty$ (see \citealp{powell:91} and \citealp[Sec 3.4]{koenker:05}). Secondly, we obtain from $Q_n(\tau)$ and Eq.\ \eqref{project} the inverse matrix of eigenvectors $T_n(\tau)$, which, by the continuous mapping theorem, is a consistent estimator. Finally, we simulate the limiting stochastic processes similarly to the case where Assumption \ref{ass:D} is met, but we substitute ${\cal B}(\cdot)$ (${\cal S}(\cdot)$) with $\widetilde{\cal B}(\cdot)$ ($\widetilde{\cal S}(\cdot)$). Clearly, this procedure becomes very time-consuming when testing at many quantiles. Nevertheless, an important exception is given for $p = 1 < k$,\footnote{Note that the case $k = 1$ is trivial because then $w = 1$ and assumption \ref{ass:D} is automatically satisfied.} where proper scaling of the test statistics ensures a pivotal limiting distribution; e.g. for a given $\tau \in \cal T$
\[
h(\tau)\frac{{\sf LM}_{2}(\tau)}{\sqrt{1+Q_n(\tau)Q_n(\tau)'}} \rightarrow_d |{\cal B}_1(1)| \equiv \sqrt{\chi^2(1)}.
\]
Unfortunately, a similar re-scaling does not work for other values $p > 1$. Therefore, we propose an additional resampling procedure that especially for the `many-$\tau$' case is significantly less time consuming.

In particular, we propose a bootstrap procedure which is inspired by \citet{rw:13} and that is valid both in the `fixed-$\tau$' case and the `many-$\tau$' case. Note, however, that we actually need the bootstrap only in the former case because the computational burden of the procedure described above for the `fixed-$\tau$' case is manageable. The algorithm for obtaining one bootstrap sample $\{(\widehat y_{i,b},x_{i,b}), 1 \leq i \leq n\}, b \in \{1,\ldots,B\}$, for a large value of $B$ is as follows:

{\bf Algorithm 1.}
\begin{enumerate}[leftmargin=40pt]
     \item[\it Step 1] Draw with replacement $\{x_{i,b}, 1 \leq i \leq n\}$ from the realized regressors $\{x_{i}, 1 \leq i \leq n\}$
      \item[\it Step 2] For each $1 \leq i \leq n$, set
      $$\widehat y_{i,b}  = \alpha_n(U_{i,b})'w_{i,b},$$
      where $\{U_{i,b}, 1 \leq i \leq n\}$ is a simulated {\sf IID} sequence of standard uniformly distributed random variables on the interval $(0,1)$, $\alpha_n$ is the restricted quantile estimator and $x_{i,b} = (z_{i,b}',w_{i,b}')'$ with the same dimensions as in the realized data.
      \item[\it Step 3] Use the bootstrap data $\{(\widehat y_{i,b},x_{i,b}), 1 \leq i \leq n\}$ to obtain bootstrap estimates $H_{n,b}(\lambda,\tau,\tilde\beta_{n,b}(\tau))$, say, of $H_{n}(\lambda,\tau,\tilde\beta_{n}(\tau))$ and construct the corresponding test statistics for $\tau \in {\cal T}$.
      \item[\it Step 4]  Our tests reject if they exceed the corresponding bootstrap critical values $\widehat c(\alpha)$, say, for some $\alpha \in (0,1).$
 \end{enumerate}

The algorithm above means that, for generating a bootstrap sample, we first draw with replacement from the regressors, where random sampling is justified by Assumption \ref{ass:A} (i.e. Corollary \ref{cor1} and Corollary \ref{cor4} are unaffected by the temporal dependence of $x_i$). The corresponding $y$-values are obtained by applying the estimated quantile function on randomly chosen standard uniformly distributed random variables. Thus, it is ensured that, under the null hypothesis, we asymptotically generate data from the distribution of $(y,x)$ with $Q_y(\tau \vert x) = x'\beta(\tau)$ so that, in {\it Step 3}, no centring of $H_{n,b}(\lambda,\tau,\tilde\beta_{n,b}(\tau))$ is needed. Here, it is crucial to draw from a uniform distribution on the whole interval $(0,1)$ in {\it Step 2} to get simulated data from the whole conditional distribution of $y$ given $x$, although the interval ${\cal T}$ is a strict subset of $(0,1)$.
So, the restriction is stronger than actually necessary, but the null distribution is still enforced. Under the alternative hypothesis, the critical values remain stochastically bounded as the validity of the null hypothesis is enforced within the generation of the bootstrap sample. 

These considerations are summarized in Proposition \ref{prop:boot}. For simplicity, consider the `many-$\tau$' case and let us generically represent our test statistics in Eq.\ \eqref{eq:GCtest1} as $\varphi({\sf LM}_1+{\sf LM2})$ using the weighting function $\varphi: [0,1]\times {\cal T} \mapsto \mathbb{R}$ from \eqref{psi}.

\begin{proposition}\label{prop:boot}\textcolor{white}{.} Let $\alpha \in (0,1)$ and assume that Assumptions \ref{ass:A}, \ref{ass:B}, and \ref{ass:C} hold uniformly in $\tau \in {\cal T}$.
    \begin{enumerate}
        \item[\textnormal{($i$)}] Under the null hypothesis 
        \[
        {\sf P}(\varphi({\sf LM}_1+{\sf LM}_2) \geq \hat{c}(\alpha)) \rightarrow \alpha.
        \]
         \item[\textnormal{($ii$)}] Under fixed alternatives 
                 \[
        {\sf P}(\varphi({\sf LM}_1+{\sf LM}_2) \geq \hat{c}(\alpha)) \rightarrow 1.
        \]
    \end{enumerate}
\end{proposition}

\subsection{Identifying Regimes of Granger Causality}

If the null hypothesis $H_0$ is rejected by the new tests, it might be interesting to get more information about the reason for the rejection. It is possible that $H_{0,1}$ is violated, i.e. that the parameters are not constant over time, and it is possible that $H_{0,2}$ is violated, i.e. that the parameters are not $0$ for all time points. To get more information, we suggest to apply an iterative procedure, in the case that $H_0$ is rejected in the first step. 

To be more specific, we follow the discussion in \cite{oka:11} and introduce to this end the scaled subgradient
\[
H_{a,b,n}(\lambda,\tau,t) \coloneqq \left(\sum_{i=\floor{an}+1}^{\floor{bn}} x_ix_i'\right)^{-1}\sum_{i=\floor{an}+1}^{\floor{\lambda n}} x_i\varphi_\tau(y_i - x_i't),
\]
defined on $\lambda \in [a,b] \subset [0,1]$, while $\tilde\beta_{a,b,n}(\tau) \coloneqq (0,\alpha_{a,b,n}(\tau)')'$ denotes the constrained estimator on the restricted sample, where
    \begin{align*} 
    \alpha_{a,b,n}(\tau) \coloneqq \argmin\limits_{\alpha \in \mathbb{R}^k} \sum_{i=\floor{a n}+1}^{\floor{bn}}\rho_{\tau}(y_i-w_{i}'\alpha).
    \end{align*}
Next, define (${sup\sf{LM}}_{a,b}$ could be defined analogously)
\begin{equation}\label{eq:GCtestab}
   \textnormal{\it{exp}\sf{LM}}_{a,b} \coloneqq \int_{\cal T}{\sf exp}\left[\frac{1}{2}\left(\ssup\limits_{\lambda \in [a,b]}{\sf LM}_{a,b,1}(\lambda,\tau) + {\sf LM}_{a,b,2}(\tau)\right)\right] d\tau,
\end{equation}
where the subsample detectors for violations of $H_{0,1}$ and $H_{1,1}$ are respectively given by
\[
{\sf LM}_{a,b,1}(\lambda,\tau) \coloneqq \left\Vert R'H_{a,b,n}(\lambda,\tau,\tilde\beta_{a,b,n}(\tau))-\lambda R'H_{a,b,n}(b ,\tau,\tilde\beta_{a,b,n}(\tau)) \right\Vert_\infty
\]
and
${\sf LM}_{a,b,2}(\tau) \coloneqq \left\Vert R'H_{a,b,n}(b,\tau,\tilde\beta_{a,b,n}(\tau))\right\Vert_\infty,$ while we define the exponentially weighted CUSUM statistic (a {\it sup}{\sf CUSUM} statistic is analogously defined)
$${\it exp}{\sf CUSUM}_{a,b}\coloneqq \ssup\limits_{\lambda \in [a,b]}\int_{\tau \in {\cal T}} {\sf exp}\left(\frac1{2}{\sf LM}_{a,b,1}(\lambda,\tau)\right)d\tau.$$ 
In the following, we describe an algorithm which identifies different regimes, in which Granger causality (GC, henceforth) holds or not. For ease of exposition, we allow to identify at most three different regimes, i.e. at most two break points. But the algorithm could straightforwardly be extended for the case of more than two breaks.

\newpage
{\small \singlespacing
{\bf Algorithm 2}.
\begin{enumerate}[leftmargin=40pt]
     \item[\it Step 1] Perform the {\it exp}{\sf LM} test at a significance of $\alpha$. If the test rejects go to Step 2, otherwise terminate. 
\item[\it Step 2] Perform the ${\it exp}{\sf CUSUM}_{0,1}$ test at significance level $1-(1-\alpha)^{\frac1{2}}$. If the test fails to reject, GC is detected on $[0,1]$; terminate. Otherwise, the first breakpoint
\(
\lambda_{1} \coloneqq \argmax\limits_{\lambda \in [0,1]}\int_{\tau \in\mathcal T} {\sf exp}\left(\frac1{2}{\sf LM}_1(\lambda,\tau)\right)d\tau
\)
is announced; go to Step 3.
\item[\it Step 3] Perform the \({\it exp}{\sf LM}_{0,\lambda_1}\) and \({\it exp}{\sf LM}_{\lambda_1,1}\) tests at the significance level $1-(1-\alpha)^{\frac1{4}}$:
\begin{itemize}[leftmargin=0pt]
    \item[1.] If both tests fail to reject, the procedure is inconclusive on $[0,1]$; terminate.
    \item[2.] If the tests fail to reject on $[0,\lambda_{1}]$ but reject on $[\lambda_{1},1]$, no GC is detected on $[0,\lambda_{1}]$. Perform the ${\it exp}{\sf CUSUM}_{\lambda_{1},1}$ test at a significance level of $1-(1-\alpha)^{\frac1{5}}$. 
    \begin{itemize}[leftmargin=20pt]
     \item[a)] If the test fails to reject, GC is detected on $[\lambda_{1},1]$.
     \item[b)] If the test rejects, then a second break $\lambda_{2} \coloneqq \argmax\limits_{\lambda \in [\lambda_{1},1]}\int_{\tau \in\mathcal T} {\sf exp}\left(\frac1{2}{\sf LM}_{\lambda_{1},1}(\lambda,\tau)\right)d\tau$ is announced. Perform the ${\it exp}{\sf LM}_{\lambda_{1},\lambda_{2}}$ and ${\it exp}{\sf LM}_{\lambda_{2},1}$ test at a significance level of $1-(1-\alpha)^{\frac1{7}}$.
         \begin{itemize}[leftmargin=20pt]
     \item[i)] If both tests fail to reject, conclude that the procedure is inconclusive on $[\lambda_{1},1]$.
      \item[ii)] If both tests reject, GC is detected on both segments.
     \item[iii)] If only one test rejects, GC is detected on the respective segment.    
     \end{itemize}
    \end{itemize}
    \item[3.] The tests reject on  $[0,\lambda_{1}]$ but fail to reject on $[\lambda_{1},1]$: See Case 2 with the roles of $[0,\lambda_{1}]$ and $[\lambda_{1},1]$ inverted.
        \item[4.] If both tests reject, compute $\mmax\limits_{1 \leq  k \leq 2}{\it exp}{\sf CUSUM}_{\lambda_{k-1},\lambda_k}$; set the significance level to $1-(1-\alpha)^{\frac1{5}}$.
        \begin{itemize}[leftmargin=20pt]
     \item[a)] If the test fails to reject, then GC is detected on $[0,\lambda_{1}]$ and $[\lambda_{1},1]$.
     \item[b)] If the test rejects, a new breakpoint $\lambda_{2}$ is announced as the respective maximizer. Suppose $\lambda_{1} \leq \lambda_2$ and compute ${\it exp}{\sf LM}_{\lambda_{1},\lambda_{2}}$ and ${\it exp}{\sf LM}_{\lambda_{2},1}$; the test on $[0,\lambda_1]$ has already rejected; set the significance level to $1-(1-\alpha)^{\frac1{7}}$.
              \begin{itemize}[leftmargin=20pt]
     \item[i)] If both tests fail to reject, the procedure is inconclusive on $[\lambda_{1},1]$.
     \item[ii)] If both tests reject, GC is detected on all segments.
     \item[iii)] If only one test rejects, GC is detected on the respective segment.
     \end{itemize}
     \end{itemize}
\end{itemize}
\end{enumerate}
}

Before proceeding, a few points are worth discussing: {\it First}, the choice of the individual significance levels at each step corresponds essentially to a Šidák correction aimed at counteracting excessive accumulation of false rejections (see also \citealp[pp. 339--340]{galeano:2017}). This allows us to identify potential regimes of Granger causality with a probability of committing a type one error bounded by $\alpha$. {\it Second}, it can be conjectured, based on the results collected in \cite{oka:11}, that the estimated break locations are $n$-consistent and their effect on the limiting distributions of the tests in {\it Step 3} will thus be asymptotically negligible. This conjecture is supported by the finite sample evidence of the next section. {\it Third}, in practice, the application of the CUSUM  statistic in {\it Step 1} can be accompanied by plotting the CUSUM curve; a point that will be illustrated in our empirical application. {\it Fourth}, regarding {\it Step 2}, note that the CUSUM test also has non-trivial power, if there is more than one break point in a given interval (see, e.g. the discussion in \citealp[Section 5]{qu:2008}). Fifth, regarding {\it Step 3.4}, note that we calculate the maximum of the CUSUM statistics in order to reduce the total number of applied tests to two; see also \citet{bomw:24} for details about these two comments. The limiting distribution of the maximum of the two CUSUM statistics is that of the maximum of two independent Brownian pillow processes and can be simulated under the null of no additional break (see \citealp[Theorem 3]{oka:11}). {\it Fifth}, and contrary to the CUSUM statistics, that are nuisance parameter free even if Assumption \ref{ass:D} fails, we suggest to apply the bootstrap procedure of Algorithm 1 in {\it Step 1} to get critical values for the various ${\it exp}{\sf LM}$ statistics in case violation of Assumption \ref{ass:D} is suspected. 


\section{Monte Carlo Simulations}
In the small sample simulations, we consider a location-scale model
\begin{align*}
y_i = \,& w_i + \gamma_i z_i + (1+\alpha w_i)\eps_i, 
\end{align*}
where $(w_1,z_1,\eps_1),\dots, (w_n,z_n,\eps_n)$ are {\sf IID} copies\footnote{Appendix B contains additional Monte Carlo simulation results based on a quantile autoregressive distributed lag model with time trends. Here, no substantial differences are observed.} of 
\[
w \sim  \chi^2(3),\quad  z, \eps \sim \mathcal{N}(0,1), \quad \text{ and }\; \eps \perp (w,z).
\]

First, we investigate the size properties setting $\gamma_i = 0$ for all $i$. We distinguish between homoscedasticity ($\alpha = 0$) and heteroscedasticity ($\alpha = 3$). In the latter case, Assumption \ref{ass:D} will only be satisfied if $\Ex[z \mid w] = \Ex[z]$. Therefore, we consider three scenarios: (1) $\alpha = 0$ {\it \&} $z \perp w$, (2) $\alpha = 3$ {\it \&} $z \perp w$, (3) $\alpha = 3$ {\it \&} ${\sf cov}[z,w] \simeq -3/4$. Thus, it is only in scenario (3) where the asymptotic critical values based on the asymptotic approximation under Assumption \ref{ass:D} are wrong. 

We perform tests at five selected quantiles $\tau \in \{0.05,0.25,0.50,0.75,0.95\}$ as well as across the complete interval $[0.05,0.95]$. We consider the case where test statistics are compared to critical values obtained under Assumption \ref{ass:D} (labelled `{\sf asy}'). Alternatively, we adjust the statistics as explained in Section 3.3 when testing at a given $\tau$ or, when testing across all $\tau \in [0,1]$, use the bootstrap  (both labelled `{\sf adj}'). For comparison, we compute also the {\it sup}{\sf Wald} test of \cite{koemach:99} given by
\begin{align}\label{supWald}
  {sup}{\sf Wald} = \ssup\limits_{\tau \in {\cal T}} nh^2(\tau)\gamma_n(\tau)'\Omega^{-1}_n(\tau)\gamma_n(\tau) \stackrel{H_{0}}{\rightarrow_d} \ssup\limits_{\tau \in {\cal T}} h^2 (\tau){\cal BB}_p(\tau)'{\cal BB}_p(\tau)  
\end{align}
where we equip the statistic with (pairs) bootstrap standard errors $\Omega_n(\cdot)$ implemented using the {\sf quantreg} package of {\sf R} (\citealp{koenker:18}). Critical values are easily obtained from a discrete approximation of the Bessel limiting process. For all bootstrap procedures we use $B = 499$ replications. All test decision are carried out at the five per cent significance level. 

Table \ref{tab1} contains the Monte Carlo results under the null hypothesis of Granger non-causality based on 2,000 Monte Carlo repetitions. As can be seen from panel $a$) and $b$) of Table \ref{tab1}, size is controlled if $n$ is moderately large irrespective of conditional homoscedasticity/heteroscedasticity because conditional mean independence and thus Assumption \ref{ass:D} is satisfied. As suggested by our theory, the performance of the tests using the asymptotic approximation derived under Assumption \ref{ass:D} deteriorates if $\cov[z,w] \neq 0$ (cf. panel $c$) of Table \ref{tab1}). However, in this case the adjustment/bootstrap alternatives do their job by effectively keeping size. The empirical size of the {\it sup}{\sf Wald} test is in all scenarios in line with the nominal significance level.   

\setlength{\tabcolsep}{2.2pt}
\begin{sidewaystable}
    \centering \footnotesize
    \begin{tabular}{lrccccccccccccccccccccccc}
    \toprule
    &   &\multicolumn{8}{c}{\sf $a$) Homosc. \& uncorr. $x$} &\multicolumn{7}{c}{\sf $b$) Heterosc. \& uncorr. $x$} &&\multicolumn{7}{c}{\sf $c$) Heterosc. \& corr. $x$} \\
    \cmidrule(l){3-9}  \cmidrule(l){11-17} \cmidrule(l){19-25}
   &   &  .05 & .025 & .50 & .75 & .95 & \multicolumn{2}{c}{[.05;.95]} & & .05 & .025 & .50 & .75 & .95 & \multicolumn{2}{c}{[.05;.95]}& & .05 & .025 & .50 & .75 & .95 & \multicolumn{2}{c}{[.05;.95]}\\ 
      \cmidrule(l){1-25}
  &   $n=$ 150 &\multicolumn{5}{c}{\sf LM}&{\it sup}{\sf LM}&{\it exp}{\sf LM}&&\multicolumn{5}{c}{\sf LM}&{\it sup}{\sf LM}&{\it exp}{\sf LM}&&\multicolumn{5}{c}{\sf LM}&{\it sup}{\sf LM}&{\it exp}{\sf LM}\\
 \cmidrule(r){3-7} \cmidrule(r){8-9}  \cmidrule(r){11-15} \cmidrule(r){16-17}    \cmidrule(r){19-23} \cmidrule(r){24-25}
    &  {adj} & .050 & .045 & .051 & .048 & .027 & .053 & .054 && .060   & .043   & .054   & .046   & .033   & .056   & .054  &&   .091 & .074 & .076 & .062 & .035 & .048 & .066 \\  
    &  {asy} & .064 & .047 & .051 & .052 & .038 & .051 & .046 &&   .068   & .043   & .055   & .047   & .039   & .053   & .050  &&   .124 & .094 & .091 & .083 & .057 &.112 &  .100 \\ 
 \cmidrule(r){3-7} \cmidrule(r){8-9}  \cmidrule(r){11-15} \cmidrule(r){16-17}  \cmidrule(r){19-23}  \cmidrule(r){24-25} 
     &  &\multicolumn{5}{c}{\sf Wald}&\multicolumn{2}{c}{{\it sup}\sf W}&&\multicolumn{5}{c}{\sf Wald}&\multicolumn{2}{c}{{\it sup}\sf W}&&\multicolumn{5}{c}{\sf Wald}&\multicolumn{2}{c}{{\it sup}\sf W}\\
          \cmidrule(r){3-7} \cmidrule(r){8-9} \cmidrule(r){11-15} \cmidrule(r){16-17} \cmidrule(r){19-23}  \cmidrule(r){24-25} 
    &   {asy}& .063 & .045 & .054 & .058 & .068 & \multicolumn{2}{c}{.067} & &        .041   & .032   & .043   & .047   & .049   &  \multicolumn{2}{c}{.043} & &        .052 & .049 & .034 &.038 & .048 &\multicolumn{2}{c}{.040} \\  \cmidrule(l){1-25}\\[-.3cm]
  &   $n=$ 300 &\multicolumn{5}{c}{\sf LM}&{\it sup}{\sf LM}&{\it exp}{\sf LM}&&\multicolumn{5}{c}{\sf LM}&{\it sup}{\sf LM}&{\it exp}{\sf LM}&&\multicolumn{5}{c}{\sf LM}&{\it sup}{\sf LM}&{\it exp}{\sf LM}\\
 \cmidrule(r){3-7} \cmidrule(r){8-9}  \cmidrule(r){11-15} \cmidrule(r){16-17}   \cmidrule(r){19-23} \cmidrule(r){24-25}
         &  {adj} & .049 & .048 & .043 & .047 & .039 & .044 & .042 && .060   & .044   & .048   & .043   & .038   & .053   & .048   && .072 & .058 & .073 & .060 & .034 & .047 & .055 \\ 
       &  {asy} & .064 & .052 & .044 & .048 & .047 & .037 & .039& &0.065   & .047   & .048   & .043   & .043   & .052   & .045  & & .103 & .090 & .093 & .090 & .067 & .115 & .094 \\  
 \cmidrule(r){3-7} \cmidrule(r){8-9}  \cmidrule(r){11-15} \cmidrule(r){16-17}  \cmidrule(r){19-23}  \cmidrule(r){24-25} 
     &  &\multicolumn{5}{c}{\sf Wald}&\multicolumn{2}{c}{{\it sup}\sf W}&&\multicolumn{5}{c}{\sf Wald}&\multicolumn{2}{c}{{\it sup}\sf W}&&\multicolumn{5}{c}{\sf Wald}&\multicolumn{2}{c}{{\it sup}\sf W}\\
          \cmidrule(r){3-7} \cmidrule(r){8-9} \cmidrule(r){11-15} \cmidrule(r){16-17} \cmidrule(r){19-23}  \cmidrule(r){24-25} 
        &   {asy}& .058 & .052 & .047 & .048 & .060 &  \multicolumn{2}{c}{.062 }&&    .043   & .039   & .041   & .046   & .039   & \multicolumn{2}{c}{.038} &&  .047 & .041 & .041 & .047 & .049 & \multicolumn{2}{c}{.039}  \\ \cmidrule(l){1-25}\\[-.3cm]
&   $n=$ 1,000 &\multicolumn{5}{c}{\sf LM}&{\it sup}{\sf LM}&{\it exp}{\sf LM}&&\multicolumn{5}{c}{\sf LM}&{\it sup}{\sf LM}&{\it exp}{\sf LM}&&\multicolumn{5}{c}{\sf LM}&{\it sup}{\sf LM}&{\it exp}{\sf LM}\\
 \cmidrule(r){3-7} \cmidrule(r){8-9}  \cmidrule(r){11-15} \cmidrule(r){16-17}   \cmidrule(r){19-23} \cmidrule(r){24-25}
       &  {adj} & .046 & .049 & .049 & .049 & .033 & .048 & .051 && .042   & .058   & .055   & .045   & .034   & .054   & .053  & &.044 & .061 & .058 & .048 & .033 & .044 & .053 \\    
        &  {asy} & .050 & .050 & .049 & .049 & .038 & .048 & .052&& .046   & .059   & .055   & .045   & .041   & .055   & .049 && .099 & .100 & .091 & .086 & .081 & .115 & .095 \\  
 \cmidrule(r){3-7} \cmidrule(r){8-9}  \cmidrule(r){11-15} \cmidrule(r){16-17}  \cmidrule(r){19-23}  \cmidrule(r){24-25} 
     &  &\multicolumn{5}{c}{\sf Wald}&\multicolumn{2}{c}{{\it sup}\sf W}&&\multicolumn{5}{c}{\sf Wald}&\multicolumn{2}{c}{{\it sup}\sf W}&&\multicolumn{5}{c}{\sf Wald}&\multicolumn{2}{c}{{\it sup}\sf W}\\
               \cmidrule(r){3-7} \cmidrule(r){8-9} \cmidrule(r){11-15} \cmidrule(r){16-17} \cmidrule(r){19-23}  \cmidrule(r){24-25} 
       &   {asy}& .051 & .047 & .050 & .051 & .046 & \multicolumn{2}{c}{.067} & & .048   & .059   & .044   & .053   & .041   & \multicolumn{2}{c}{.058} &&  .057 & .052 & .047 & .046 & .048 & \multicolumn{2}{c}{.055}\\ \cmidrule(l){1-25}\\[-.3cm]
&   $n=$ 2,000 &\multicolumn{5}{c}{\sf LM}&{\it sup}{\sf LM}&{\it exp}{\sf LM}&&\multicolumn{5}{c}{\sf LM}&{\it sup}{\sf LM}&{\it exp}{\sf LM}&&\multicolumn{5}{c}{\sf LM}&{\it sup}{\sf LM}&{\it exp}{\sf LM}\\
 \cmidrule(r){3-7} \cmidrule(r){8-9}  \cmidrule(r){11-15} \cmidrule(r){16-17}   \cmidrule(r){19-23} \cmidrule(r){24-25}
         &  {adj} & .050 & .054 & .054 & .057 & .044 & .046 & .049 && .049   & .056   & .054   & .050   & .041   & .044   & .050   && .033 & .052 & .050 & .047 & .026 & .045 & .051 \\  
         &  {asy} & .053 & .054 & .054 & .057 & .046 & .042 & .052 &&0.054   & .056   & .054   & .050   & .047   & .045   & .052    && .097 & .095 & .089 & .094 & .082 & .117 & .091 \\ 
 \cmidrule(r){3-7} \cmidrule(r){8-9}  \cmidrule(r){11-15} \cmidrule(r){16-17}  \cmidrule(r){19-23}  \cmidrule(r){24-25} 
  &  &\multicolumn{5}{c}{\sf Wald}&\multicolumn{2}{c}{{\it sup}\sf W}&&\multicolumn{5}{c}{\sf Wald}&\multicolumn{2}{c}{{\it sup}\sf W}&&\multicolumn{5}{c}{\sf Wald}&\multicolumn{2}{c}{{\it sup}\sf W}\\
               \cmidrule(r){3-7} \cmidrule(r){8-9} \cmidrule(r){11-15} \cmidrule(r){16-17} \cmidrule(r){19-23}  \cmidrule(r){24-25} 
        &   {asy}& .058 & .052 & .051 & .050 & .049 & \multicolumn{2}{c}{.063} && .050   & .042   & .051   & .052   & .047   &\multicolumn{2}{c}{.060} && .046 & .050 & .047 & .050 & .051 & \multicolumn{2}{c}{.055} \\  \bottomrule
    \end{tabular}
    \caption{Rejection frequencies at a nominal size of five per cent of various tests statistics constructed over a grid $[0.05,0.06,\dots,0.95]$ based on 2,000 Monte Carlo repetitions.}\label{tab1}
\end{sidewaystable}

\normalsize

Turning to the power properties, we set, for better comparison,  $\alpha = 0$ and $z \perp w$ (i.e. Assumption \ref{ass:D} is satisfied) and evaluate tests over the complete quantile interval $[0.05,0.95]$. First, let us consider the following three break scenarios:
\[
\textnormal{\sf A}:\; \gamma_i = \begin{dcases}
 \gamma & i \leq \floor{n/2} \\
-\gamma & \text{otherwise} 
\end{dcases}
\qquad
\textnormal{\sf B}:\; \gamma_i = \begin{dcases}
0& i \leq \floor{n/2} \\
\gamma & \text{otherwise} 
\end{dcases}
\qquad
\textnormal{\sf C}: \gamma_{i}=\gamma.
\]
This means that, in Scenario {\sf A}, we have a structural break in the Granger parameter and the parameters sum up to zero over time (i.e. $\int_0^1 g(v){\sf d}v = 0$). Also in Scenario {\sf B}, there is a structural break, but the sum over time is not zero. In Scenario {\sf C}, the Granger parameter is constant and equal to $\gamma$ (i.e. $\Delta(\tau) \neq 0$).

\begin{figure}[!ht]
\def\dist{0.35}
\def\dista{-1cm}
\def\distb{-.25cm}
\hspace*{\distb}
\centering
\begin{subfigure}[b]{\dist\textwidth}
\centering 
   \begin{tikzpicture}
        \begin{axis}[
        xmin = 0, xmax = 0.3, ymin = 0,ymax=1,	
                yticklabel style={
        /pgf/number format/fixed,
        /pgf/number format/precision=5
}, 
xticklabel style={
        /pgf/number format/fixed,
        /pgf/number format/precision=5
},
scaled y ticks=false,
tick label style={font=\scriptsize},
		label style={font=\small},
		tick label style={font=\scriptsize},
		legend style={font=\scriptsize},
  width=1.075\textwidth,
    height=.25\textheight,
			y label style={at={(axis description cs:0.02,.5)},anchor=north},
xticklabels=\empty,  
  clip=false,
  legend style={nodes={scale=0.65, transform shape}, draw=none,at={(1.08,0.8)},anchor=north},legend cell align={left} 
        ]
        \draw[black!30,thick] (axis cs:\pgfkeysvalueof{/pgfplots/xmin},0.05) -- (axis cs:\pgfkeysvalueof{/pgfplots/xmax},0.05);
               \draw [font=\scriptsize] (0,95) node [right] {A $n=$ 150};
                      \addplot[solid] table [x=tau,y=C]{plot/power150.txt};
\addplot[orange,solid] table [x=tau,y=A]{plot/power150.txt};
\addplot[blue,solid] table [x=tau,y=B]{plot/power150.txt};
         \end{axis}
   \end{tikzpicture}
\end{subfigure}  \hspace*{\dista}
\begin{subfigure}[b]{\dist\textwidth}
 \centering 
    \begin{tikzpicture}
        \begin{axis}[
        xmin = 0, xmax = 0.3, ymin = 0,ymax=1,	
                yticklabel style={
        /pgf/number format/fixed,
        /pgf/number format/precision=5
}, 
xticklabel style={
        /pgf/number format/fixed,
        /pgf/number format/precision=5
},
scaled y ticks=false,
tick label style={font=\scriptsize},
		label style={font=\small},
		tick label style={font=\scriptsize},
		legend style={font=\scriptsize},
  width=1.075\textwidth,
    height=.25\textheight,
			y label style={at={(axis description cs:0.02,.5)},anchor=north},
xticklabels=\empty,  
yticklabels=\empty, 
  clip=false,
  legend style={nodes={scale=0.65, transform shape}, draw=none,at={(1.08,0.8)},anchor=north},legend cell align={left} 
        ]
        \draw[black!30,thick] (axis cs:\pgfkeysvalueof{/pgfplots/xmin},0.05) -- (axis cs:\pgfkeysvalueof{/pgfplots/xmax},0.05);
               \draw [font=\scriptsize] (0,95) node [right] {B $n=$ 150};
                      \addplot[solid] table [x=tau,y=C]{plot/power150B.txt};
\addplot[orange,solid] table [x=tau,y=A]{plot/power150B.txt};
\addplot[blue,solid] table [x=tau,y=B]{plot/power150B.txt};
         \end{axis}
   \end{tikzpicture}
\end{subfigure}\hspace*{\dista}
\begin{subfigure}[b]{\dist\textwidth}
 \centering 
    \begin{tikzpicture}
        \begin{axis}[
        xmin = 0, xmax = 0.3, ymin = 0,ymax=1,	
                yticklabel style={
        /pgf/number format/fixed,
        /pgf/number format/precision=5
}, 
xticklabel style={
        /pgf/number format/fixed,
        /pgf/number format/precision=5
},
scaled y ticks=false,
tick label style={font=\scriptsize},
		label style={font=\small},
		tick label style={font=\scriptsize},
		legend style={font=\scriptsize},
  width=1.075\textwidth,
    height=.25\textheight,
			y label style={at={(axis description cs:0.02,.5)},anchor=north},
xticklabels=\empty,  
yticklabels=\empty, 
  clip=false,
  legend style={nodes={scale=0.65, transform shape}, draw=none,at={(1.08,0.8)},anchor=north},legend cell align={left} 
        ]
        \draw[black!30,thick] (axis cs:\pgfkeysvalueof{/pgfplots/xmin},0.05) -- (axis cs:\pgfkeysvalueof{/pgfplots/xmax},0.05);
               \draw [font=\scriptsize] (0,95) node [right] {C $n=$ 150};
                      \addplot[solid] table [x=tau,y=C]{plot/power150C.txt};
\addplot[orange,solid] table [x=tau,y=A]{plot/power150C.txt};
\addplot[blue,solid] table [x=tau,y=B]{plot/power150C.txt};
         \end{axis}
   \end{tikzpicture}
\end{subfigure}
\hspace*{\distb}
\begin{subfigure}[b]{\dist\textwidth} 
\centering
   \begin{tikzpicture}
        \begin{axis}[
        xmin = 0, xmax = 0.3, ymin = 0,ymax=1,	
                yticklabel style={
        /pgf/number format/fixed,
        /pgf/number format/precision=5
}, 
xticklabel style={
        /pgf/number format/fixed,
        /pgf/number format/precision=5
},
scaled y ticks=false,
tick label style={font=\scriptsize},
		label style={font=\small},
		tick label style={font=\scriptsize},
		legend style={font=\scriptsize},
  width=1.075\textwidth,
    height=.25\textheight,
			y label style={at={(axis description cs:0.02,.5)},anchor=north},
xticklabels=\empty,  
  clip=false,
  legend style={nodes={scale=0.65, transform shape}, draw=none,at={(1.08,0.8)},anchor=north},legend cell align={left} 
        ]
        \draw[black!30,thick] (axis cs:\pgfkeysvalueof{/pgfplots/xmin},0.05) -- (axis cs:\pgfkeysvalueof{/pgfplots/xmax},0.05);
               \draw [font=\scriptsize] (0,95) node [right] {A $n=$ 300};
                      \addplot[solid] table [x=tau,y=C]{plot/power300.txt};
\addplot[orange,solid] table [x=tau,y=A]{plot/power300.txt};
\addplot[blue,solid] table [x=tau,y=B]{plot/power300.txt};
         \end{axis}
   \end{tikzpicture} 
\end{subfigure} \hspace*{\dista}
\begin{subfigure}[b]{\dist\textwidth} 
\centering
   \begin{tikzpicture}
        \begin{axis}[
        xmin = 0, xmax = 0.3, ymin = 0,ymax=1,	
                yticklabel style={
        /pgf/number format/fixed,
        /pgf/number format/precision=5
}, 
xticklabel style={
        /pgf/number format/fixed,
        /pgf/number format/precision=5
},
scaled y ticks=false,
tick label style={font=\scriptsize},
		label style={font=\small},
		tick label style={font=\scriptsize},
		legend style={font=\scriptsize},
  width=1.075\textwidth,
    height=.25\textheight,
			y label style={at={(axis description cs:0.02,.5)},anchor=north},
xticklabels=\empty,  
yticklabels=\empty, 
  clip=false,
  legend style={nodes={scale=0.65, transform shape}, draw=none,at={(1.08,0.8)},anchor=north},legend cell align={left} 
        ]
        \draw[black!30,thick] (axis cs:\pgfkeysvalueof{/pgfplots/xmin},0.05) -- (axis cs:\pgfkeysvalueof{/pgfplots/xmax},0.05);
               \draw [font=\scriptsize] (0,95) node [right] {Bs $n=$ 300};
                      \addplot[solid] table [x=tau,y=C]{plot/power300B.txt};
\addplot[orange,solid] table [x=tau,y=A]{plot/power300B.txt};
\addplot[blue,solid] table [x=tau,y=B]{plot/power300B.txt};
         \end{axis}
   \end{tikzpicture}
\end{subfigure}\hspace*{\dista}
\begin{subfigure}[b]{\dist\textwidth} 
\centering
   \begin{tikzpicture}
        \begin{axis}[
        xmin = 0, xmax = 0.3, ymin = 0,ymax=1,	
                yticklabel style={
        /pgf/number format/fixed,
        /pgf/number format/precision=5
}, 
xticklabel style={
        /pgf/number format/fixed,
        /pgf/number format/precision=5
},
scaled y ticks=false,
tick label style={font=\scriptsize},
		label style={font=\small},
		tick label style={font=\scriptsize},
		legend style={font=\scriptsize},
  width=1.075\textwidth,
    height=.25\textheight,
			y label style={at={(axis description cs:0.02,.5)},anchor=north},
xticklabels=\empty,  
yticklabels=\empty,  
  clip=false,
  legend style={nodes={scale=0.65, transform shape}, draw=none,at={(1.08,0.8)},anchor=north},legend cell align={left} 
        ]
        \draw[black!30,thick] (axis cs:\pgfkeysvalueof{/pgfplots/xmin},0.05) -- (axis cs:\pgfkeysvalueof{/pgfplots/xmax},0.05);
               \draw [font=\scriptsize] (0,95) node [right] {C $n=$ 300};
                      \addplot[solid] table [x=tau,y=C]{plot/power300C.txt};
\addplot[orange,solid] table [x=tau,y=A]{plot/power300C.txt};
\addplot[blue,solid] table [x=tau,y=B]{plot/power300C.txt};
         \end{axis}
   \end{tikzpicture}
\end{subfigure}
\hspace*{-0cm}
\begin{subfigure}[t]{\dist\textwidth} 
\centering
   \begin{tikzpicture}
        \begin{axis}[
        xmin = 0, xmax = 0.3, ymin = 0,ymax=1,	
                yticklabel style={
        /pgf/number format/fixed,
        /pgf/number format/precision=5
}, 
xticklabel style={
        /pgf/number format/fixed,
        /pgf/number format/precision=5
},
scaled y ticks=false,
tick label style={font=\scriptsize},
		label style={font=\small},
		tick label style={font=\scriptsize},
		legend style={font=\scriptsize},
  width=1.075\textwidth,
    height=.25\textheight,
			y label style={at={(axis description cs:0.02,.5)},anchor=north},
  clip=false,
  legend style={nodes={scale=0.65, transform shape}, draw=none,at={(1.08,0.8)},anchor=north},legend cell align={left} 
        ]
        \draw[black!30,thick] (axis cs:\pgfkeysvalueof{/pgfplots/xmin},0.05) -- (axis cs:\pgfkeysvalueof{/pgfplots/xmax},0.05);
               \draw [font=\scriptsize] (0,95) node [right] {A $n=$ 500};
                      \addplot[solid] table [x=tau,y=C]{plot/power500.txt};
\addplot[orange,solid] table [x=tau,y=A]{plot/power500.txt};
\addplot[blue,solid] table [x=tau,y=B]{plot/power500.txt};
         \end{axis}
   \end{tikzpicture} 
\end{subfigure} \hspace*{\dista}
\begin{subfigure}[t]{\dist\textwidth} 
\centering
   \begin{tikzpicture}
        \begin{axis}[
        xmin = 0, xmax = 0.3, ymin = 0,ymax=1,	
                yticklabel style={
        /pgf/number format/fixed,
        /pgf/number format/precision=5
}, 
xticklabel style={
        /pgf/number format/fixed,
        /pgf/number format/precision=5
},
scaled y ticks=false,
tick label style={font=\scriptsize},
		label style={font=\small},
		tick label style={font=\scriptsize},
		legend style={font=\scriptsize},
  width=1.075\textwidth,
    height=.25\textheight,
			y label style={at={(axis description cs:0.02,.5)},anchor=north},
yticklabels=\empty,  
  clip=false,
  legend style={nodes={scale=0.65, transform shape}, draw=none,at={(1.08,0.8)},anchor=north},legend cell align={left} 
        ]
        \draw[black!30,thick] (axis cs:\pgfkeysvalueof{/pgfplots/xmin},0.05) -- (axis cs:\pgfkeysvalueof{/pgfplots/xmax},0.05);
               \draw [font=\scriptsize] (0,95) node [right] {B $n=$ 500};
                      \addplot[solid] table [x=tau,y=C]{plot/power500B.txt};
\addplot[orange,solid] table [x=tau,y=A]{plot/power500B.txt};
\addplot[blue,solid] table [x=tau,y=B]{plot/power500B.txt};
         \end{axis}
   \end{tikzpicture}
\end{subfigure}\hspace*{\dista}
\begin{subfigure}[t]{\dist\textwidth} 
\centering
   \begin{tikzpicture}
        \begin{axis}[
        xmin = 0, xmax = 0.3, ymin = 0,ymax=1,	
                yticklabel style={
        /pgf/number format/fixed,
        /pgf/number format/precision=5
}, 
xticklabel style={
        /pgf/number format/fixed,
        /pgf/number format/precision=5
},
scaled y ticks=false,
tick label style={font=\scriptsize},
		label style={font=\small},
		tick label style={font=\scriptsize},
		legend style={font=\scriptsize},
  width=1.075\textwidth,
    height=.25\textheight,
			y label style={at={(axis description cs:0.02,.5)},anchor=north},
yticklabels=\empty,  
  clip=false,
  legend style={nodes={scale=0.65, transform shape}, draw=none,at={(1.08,0.8)},anchor=north},legend cell align={left} 
        ]
        \draw[black!30,thick] (axis cs:\pgfkeysvalueof{/pgfplots/xmin},0.05) -- (axis cs:\pgfkeysvalueof{/pgfplots/xmax},0.05);
               \draw [font=\scriptsize] (0,95) node [right] {C $n=$ 500};
                      \addplot[solid] table [x=tau,y=C]{plot/power500C.txt};
\addplot[orange,solid] table [x=tau,y=A]{plot/power500C.txt};
\addplot[blue,solid] table [x=tau,y=B]{plot/power500C.txt};
         \end{axis}
   \end{tikzpicture}
\end{subfigure}
\vspace*{-.25cm}
             \caption{Rejection frequencies under the alternative scenarios A, B, C as a function of $\gamma \in [0,0.3]$ for $\textnormal{\it sup\sf LM}$ (orange), $\textnormal{\it exp\sf LM}$ (blue), and ${ sup\sf W}$ (black), with $B = 499$ using 2,000 Monte Carlo iterations over a grid $\tau \in [0.05,0.06,\dots,0.95]$.}\label{fig3}
\end{figure}
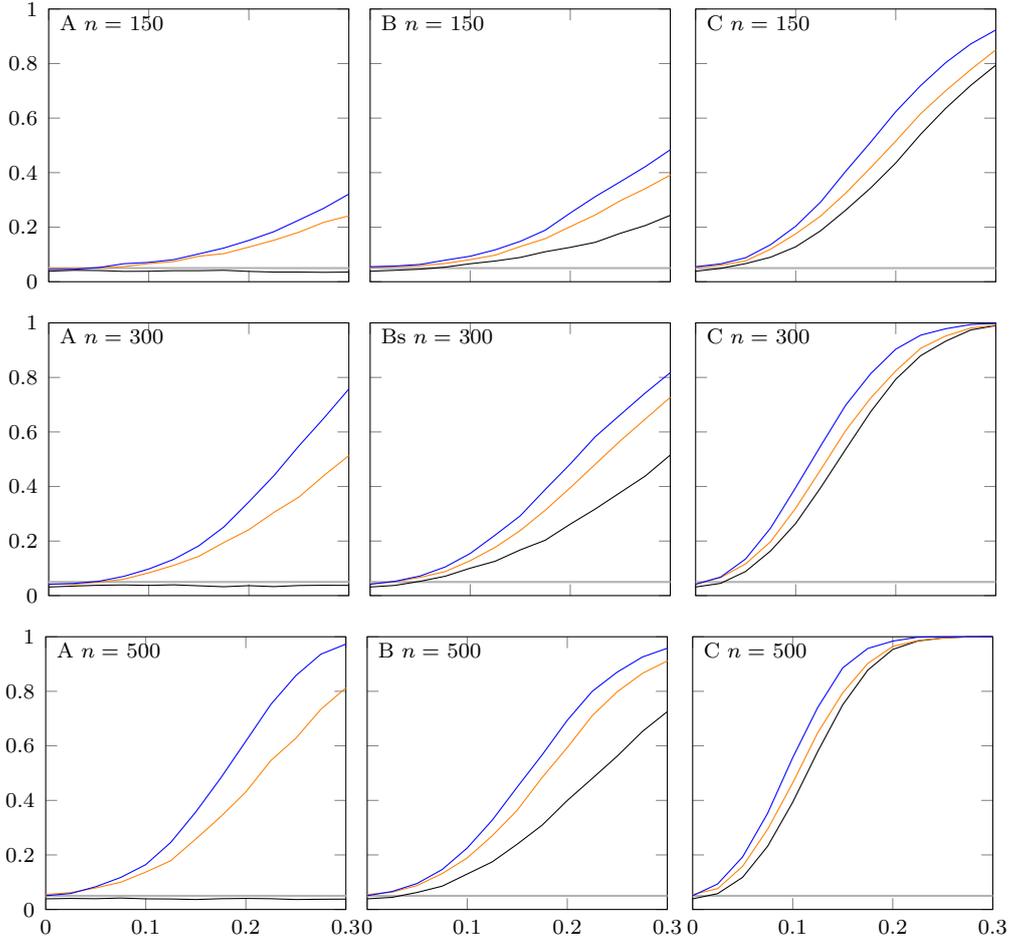

 We consider our two statistics ${\it sup}{\sf LM}$, ${\it exp}{\sf LM}$ (bootstrapped version) and the {\it sup}{\sf Wald} test. It is expected that our tests have power against all alternatives, which increases in $n$, while {\it sup}{\sf Wald} has no power in Scenario {\sf A}, also for large $n$ because $\int_0^1 g(v){\sf d}v = 0$. This is indeed the empirical result. Somewhat surprisingly, the power curves in Figure \ref{fig3} suggest that in all scenarios, the break-robust tests are more powerful than the {\it sup}{\sf Wald}; in particular, also in Scenario {\sf C}, where no break is present. Among our new tests, the {\it exp}{\sf LM} test has more power than the {\it sup}{\sf LM} test.

As a final analysis, we implement Algorithm 2 in order to identify regimes that are potentially subject to Granger causality. Due to its superior performance (see the discussion surrounding Figure \ref{fig3}), we use ${\it exp}{\sf LM}_{\lambda_j,\lambda_{j+1}}$ in conjunction with the CUSUM statistic. For this, we consider up to three regimes of the (normalized) sample $[0,1]$, i.e. $[\lambda_j,\lambda_{j+1}]$, $0 \leq j \leq 2$, with $\lambda_0 = 0$, $\lambda_3 = 1$, with $1/3 = \lambda_1 < \lambda_2 = 2/3$. We distinguish between four different scenarios: Two cases with one break $(0,0,1/2)$, $(1/2,0,0)$, and two cases with two breaks $(0,1/2,0)$, $(1/2,0,1/2)$,  where $(a,b,c)$ means that $\gamma_i$, $i = 1,\dots,n$, takes the value $a$, $b$, and/or $c$ in regime 1, 2, and 3, respectively. As can be seen from Table \ref{tab:alg}, Algorithm 2 appears to be able to consistently date the breaks and to correctly identify the regimes of Granger causality with a probability of committing a type-one error not larger than five percent. Once again, the the exponentially weighted test appears to have a superior performance. Consistent with earlier results in the literature on change point detection (see, e.g. \citealp{wied:2012}), correct detection rates are lower in cases of more than just one break.
 
\setlength{\tabcolsep}{4.66pt}
\begin{table}[h]
\begin{center} 
\begin{adjustbox}{max width=\textwidth}
    \begin{tabular}{lrccccccccccccccccccc}
    \toprule
    & &  &  \multicolumn{5}{c}{ \% \sf rejection frequencies}  \\
    \cmidrule(r){4-8}  \\[-15pt]
    &&&&& \multicolumn{3}{c}{{\sf given} {\sf LM} {\it \&} {\sf CUSUM reject}}&&\multicolumn{4}{c}{\% \sf detected}& &\multicolumn{3}{c}{$\lambda_{1,n}$}&&\multicolumn{3}{c}{$\lambda_{2,n}$}\\
    \cmidrule(r){6-8} \cmidrule(r){10-13} \cmidrule(l){15-17} \cmidrule(l){19-21} 
&   $n$  & &  {\sf LM} &  {\sf CUSUM} &  $[\lambda_0,\lambda_1]$ & $[\lambda_1,\lambda_2]$ & $[\lambda_2,\lambda_3]$ && = & < & > &  $?$& & {\sf mean} & {\sf median} &{\sf var} & & {\sf mean} & {\sf median} & {\sf var} \\ 
      \hline
\multirow{8}{*}{$(0,0,\gamma)$}&500&	{\sf sup}	&	97.30	&	90.35	&	1.94	&	100.00	&		&	&	87.75	&	11.70	&	0.55	&	0.00	&	&	0.6253	&	0.6480	&	0.0048	&	&		&		&		\\
&&	{\sf exp}	&	99.45	&	96.55	&	1.52	&	100.00	&		&	&	95.70	&	3.80	&	0.50	&	0.00	&	&	0.6406	&	0.6580	&	0.0032	&	&		&		&		\\
&1,000&	{\sf sup}	&	100.00	&	99.85	&	1.83	&	100.00	&		&	&	98.65	&	0.15	&	1.20	&	0.00	&	&	0.6428	&	0.6570	&	0.0019	&	&		&		&		\\
&&	{\sf exp}	&	100.00	&	99.95	&	1.46	&	100.00	&		&	&	99.25	&	0.05	&	0.70	&	0.00	&	&	0.6511	&	0.6610	&	0.0011	&	&		&		&		\\
&2,000&	{\sf sup}	&	100.00	&	100.00	&	1.38	&	100.00	&		&	&	98.05	&	0.00	&	1.95	&	0.00	&	&	0.6546	&	0.6625	&	0.0005	&	&		&		&		\\
&&	{\sf exp}	&	100.00	&	100.00	&	1.26	&	100.00	&		&	&	98.90	&	0.00	&	1.10	&	0.00	&	&	0.6589	&	0.6645	&	0.0003	&	&		&		&		\\
&4,000&	{\sf sup}	&	100.00	&	100.00	&	1.37	&	100.00	&		&	&	98.25	&	0.00	&	1.75	&	0.00	&	&	0.6601	&	0.6643	&	0.0002	&	&		&		&		\\
&&	{\sf exp}	&	100.00	&	100.00	&	1.22	&	100.00	&		&	&	98.50	&	0.00	&	1.50	&	0.00	&	&	0.6623	&	0.6653	&	0.0001	&	&		&		&		\\
  \cmidrule(r){4-7} \cmidrule(r){10-13} \cmidrule(l){15-17} 
\multirow{8}{*}{$(\gamma,0,0)$}&500&	{\sf sup}	&	97.60	&	90.85	&	100.00	&	1.82	&		&	&	88.20	&	11.05	&	0.75	&	0.05	&	&	0.3761	&	0.3520	&	0.0052	&	&		&		&		\\
&&	{\sf exp}	&	99.35	&	97.05	&	100.00	&	1.25	&		&	&	95.95	&	3.45	&	0.60	&	0.05	&	&	0.3582	&	0.3400	&	0.0032	&	&		&		&		\\
&1,000&	{\sf sup}	&	100.00	&	99.95	&	100.00	&	1.67	&		&	&	98.65	&	0.05	&	1.30	&	0.00	&	&	0.3583	&	0.3420	&	0.0022	&	&		&		&		\\
&&	{\sf exp}	&	100.00	&	100.00	&	100.00	&	1.56	&		&	&	99.10	&	0.00	&	0.90	&	0.00	&	&	0.3474	&	0.3370	&	0.0012	&	&		&		&		\\
&2,000&	{\sf sup}	&	100.00	&	100.00	&	100.00	&	1.88	&		&	&	98.30	&	0.00	&	1.70	&	0.00	&	&	0.3457	&	0.3370	&	0.0006	&	&		&		&		\\
&&	{\sf exp}	&	100.00	&	100.00	&	100.00	&	1.31	&		&	&	99.35	&	0.00	&	0.65	&	0.00	&	&	0.3412	&	0.3355	&	0.0003	&	&		&		&		\\
&4,000&	{\sf sup}	&	100.00	&	100.00	&	100.00	&	1.58	&		&	&	98.05	&	0.00	&	1.95	&	0.00	&	&	0.3400	&	0.3355	&	0.0002	&	&		&		&		\\
&&	{\sf exp}	&	100.00	&	100.00	&	100.00	&	1.31	&		&	&	98.95	&	0.00	&	1.05	&	0.00	&	&	0.3373	&	0.3343	&	0.0001	&	&		&		&		\\
     \cmidrule(l){2-21} 
\multirow{8}{*}{$(\gamma,0,\gamma)$}&500&	{\sf sup}	&	100.00	&	30.10	&	100.00	&	2.61	&	100.00	&	&	19.15	&	80.85	&	0.00	&	0.00	&	&	0.3331	&	0.3320	&	0.0010	&	&	0.6649	&	0.6660	&	0.0012	\\
&&	{\sf exp}	&	100.00	&	50.80	&	100.00	&	2.16	&	100.00	&	&	41.60	&	58.40	&	0.00	&	0.00	&	&	0.3288	&	0.3300	&	0.0012	&	&	0.6731	&	0.6680	&	0.0013	\\
&1,000&	{\sf sup}	&	100.00	&	75.80	&	100.00	&	1.57	&	100.00	&	&	73.40	&	26.60	&	0.00	&	0.00	&	&	0.3320	&	0.3330	&	0.0007	&	&	0.6668	&	0.6660	&	0.0006	\\
&&	{\sf exp}	&	100.00	&	93.80	&	100.00	&	1.29	&	100.00	&	&	93.25	&	6.75	&	0.00	&	0.00	&	&	0.3284	&	0.3320	&	0.0006	&	&	0.6702	&	0.6670	&	0.0006	\\
&2,000&	{\sf sup}	&	100.00	&	99.40	&	100.00	&	2.16	&	100.00	&	&	99.40	&	0.60	&	0.00	&	0.00	&	&	0.3313	&	0.3330	&	0.0003	&	&	0.6683	&	0.6665	&	0.0003	\\
&&	{\sf exp}	&	100.00	&	100.00	&	100.00	&	1.20	&	100.00	&	&	100.00	&	0.00	&	0.00	&	0.00	&	&	0.3298	&	0.3320	&	0.0002	&	&	0.6695	&	0.6675	&	0.0002	\\
&4,000&	{\sf sup}	&	100.00	&	100.00	&	100.00	&	1.40	&	100.00	&	&	100.00	&	0.00	&	0.00	&	0.00	&	&	0.3322	&	0.3333	&	0.0001	&	&	0.6679	&	0.6668	&	0.0001	\\
&&	{\sf exp}	&	100.00	&	100.00	&	100.00	&	1.05	&	100.00	&	&	100.00	&	0.00	&	0.00	&	0.00	&	&	0.3316	&	0.3330	&	0.0001	&	&	0.6685	&	0.6670	&	0.0001	\\
 \cmidrule(r){4-8} \cmidrule(r){10-13} \cmidrule(l){15-17} \cmidrule(l){19-21} 
\multirow{8}{*}{$(0,\gamma,0)$}&500&	{\sf sup}	&	90.90	&	34.40	&	2.61	&	100.00	&	4.58	&	&	22.95	&	77.05	&	0.00	&	0.00	&	&	0.3260	&	0.3300	&	0.0012	&	&	0.6684	&	0.6680	&	0.0012	\\
&&	{\sf exp}	&	96.20	&	53.10	&	1.58	&	100.00	&	3.04	&	&	44.45	&	55.55	&	0.00	&	0.00	&	&	0.3305	&	0.3320	&	0.0011	&	&	0.6679	&	0.6660	&	0.0012	\\
&1,000&	{\sf sup}	&	99.95	&	78.45	&	1.81	&	100.00	&	1.94	&	&	77.25	&	22.75	&	0.00	&	0.00	&	&	0.3294	&	0.3320	&	0.0007	&	&	0.6709	&	0.6670	&	0.0007	\\
&&	{\sf exp}	&	100.00	&	93.95	&	1.23	&	100.00	&	1.18	&	&	93.55	&	6.45	&	0.00	&	0.00	&	&	0.3299	&	0.3330	&	0.0006	&	&	0.6698	&	0.6670	&	0.0005	\\
&2,000&	{\sf sup}	&	100.00	&	99.90	&	1.65	&	100.00	&	1.65	&	&	99.90	&	0.10	&	0.00	&	0.00	&	&	0.3299	&	0.3320	&	0.0003	&	&	0.6694	&	0.6670	&	0.0003	\\
&&	{\sf exp}	&	100.00	&	100.00	&	1.20	&	100.00	&	1.75	&	&	100.00	&	0.00	&	0.00	&	0.00	&	&	0.3310	&	0.3325	&	0.0002	&	&	0.6686	&	0.6670	&	0.0002	\\
&4,000&	{\sf sup}	&	100.00	&	100.00	&	1.70	&	100.00	&	1.60	&	&	100.00	&	0.00	&	0.00	&	0.00	&	&	0.3307	&	0.3325	&	0.0001	&	&	0.6686	&	0.6668	&	0.0001	\\
&&	{\sf exp}	&	100.00	&	100.00	&	1.55	&	100.00	&	1.45	&	&	100.00	&	0.00	&	0.00	&	0.00	&	&	0.3316	&	0.3329	&	0.0001	&	&	0.6678	&	0.6668	&	0.0001	\\
    \bottomrule
    \end{tabular}
    \end{adjustbox}
    \caption{The various tests statistics constructed over a grid $[0.05,0.06,\dots,0.95]$ of quantile levels; the initial significance level is $\alpha = 0.05$; 2,000 Monte Carlo are used. Rejection frequencies in columns labelled `{\sf LM}' and `{\sf CUSUM}' are relative to total number of repetitions. The columns labelled `\% {\sf detected}' give the rate of detected breaks relative to the total number repetitions while labelled `$=$',  `$</>$', and `?' gives the rate of correctly detected breaks, too few/many detected breaks, and the rate of
 inconclusive test decision, respectively. }\label{tab:alg}
\end{center}
\end{table}

\section{Empirical Illustration: Crude Oil and Returns}

We revisit an analysis about the causal relationships between crude oil and stock returns from \citet{dingetal:16} to illustrate the advantages of our new test. \citet{dingetal:16} consider the daily returns of West Texas Intermediate ({\sf WTI}) and Dubai crude oil as well as five major (mainly Asian) stock index returns, S\&P 500 ({\sf SNP}), Nikkei ({\sf NIK}), Hang Seng ({\sf HAN}), Shanghai ({\sf SHA}), and KOSPI ({\sf KOS}), from January 1, 1996, to October 12, 2012.

Following \cite{dingetal:16}, we consider the following autoregressive distributed lag models
\begin{align*}
Q_{{\sf ROIL}_i}(\tau \mid {\cal F}_{i-1}) = \,& \beta_1(\tau) + \sum_{j=1}^{q} \alpha_{1,j}(\tau) {\sf ROIL}_{i-j} + \sum_{j=1}^{q} \gamma_{1,j}(\tau) {\sf RS}_{i-j} \\
Q_{{\sf RS}_i}(\tau \mid {\cal F}_{i-1}) = \,& \beta_2(\tau) + \sum_{j=1}^{q} \alpha_{2,j}(\tau) {\sf RS}_{i-j} + \sum_{j=1}^{q} \gamma_{2,j}(\tau) {\sf ROIL}_{i-j}
\end{align*}
for ${\cal F}_i \coloneqq \sigma(\{{\sf ROIL}_{j},{\sf RS}_{j}, j \leq i\}).$ Here, ${\sf ROIL}_i$ and ${\sf RS}_i$ denote the oil and stock returns at time point $i$, respectively, $Q_{Y_i}(\cdot \mid \mathcal{F}_{i-1})$ denotes the conditional quantile functions of a given variable $Y_i$. Of main interest are the coefficients $\gamma_{k,j}(\tau)$, which describe if one series helps to predict the conditional quantile of the other.

One of their main findings is a considerable Granger influence in quantiles of the {\sf WTI} returns on the stock returns, which is much stronger compared with the other direction (stocks on {\sf WTI}). The application of our test mainly supports this finding and, in particular, gives substantially stronger evidence for this kind of relationship. Moreover, by the use of Algorithm 2, we are able to give more insight on the timing of violations of no Granger causality. We follow their discussion (\citealp[Section 4.3]{dingetal:16}), where quantile ranges on $[0.05,0.95]$ corresponding to the complete interval, the lower (i.e. $[0.05,0.2]$) and upper tails (i.e. $[0.80,0.95]$), as well as the range around the median (i.e. $[0.40,0.60]$) are considered. 

Table \ref{TableApp1} presents the $p$-values of our bootstrapped {\it exp}{\sf LM} and {\it sup}{\sf LM} tests, as well as the {\it sup}{\sf Wald} test in Eq.\ \eqref{supWald} equipped with bootstrap standard errors, each computed over different quantile intervals using a step size of $0.01$ for the grids. In simulations, these three bootstrap-based tests display best size and power properties. In particular, the latter test equipped with bootstrap standard errors outperforms in our simulations the test equipped with a kernel based plug-in estimator for the asymptotic covariance matrix used by \cite{dingetal:16}, so that we refrain from using that test. As number of lags, we consider the values $q \in \{1,3,6,9\}$.\footnote{The results do not change substantially when using the sequential lag-length procedure of \citet{dingetal:16}, where we consider the highest number of lags (with 9 as the maximum number) for which the tests reject the null hypothesis.} 

Out of 160 scenarios in total, the $p$-values of our test are smaller than or equal to these in \citet{dingetal:16} in $160-17=143$ times. The cases in which our $p$-values are higher, mostly concern the index KOSPI, where our $p$-values are larger seven out of 16 times. But this mainly concerns cases, where the $p$-values are large anyway. So, we have a robust finding that, if there is some evidence for Granger causality, our test strengthens this evidence.

Having identified models, for which the null hypothesis is rejected, we aim at identifying regimes of Granger causality with the multi-step procedure from Algorithm 2. In order to limit computational complexity, we only consider the interval $[0.05,0,95]$ and the same lag lengths as in the first step. Table \ref{emp:regimes} yields the results for all stocks but {\sf SHA} that did not display evidence of Granger causality.

\begin{table}[!ht]
\begin{center}
\begin{adjustbox}{max width=\textwidth}
\begin{tabular}{llcccccccccccccccccc}\toprule
 & &\multicolumn{3}{c}{[.05,.95]} &&\multicolumn{3}{c}{[.05,.20]}&& \multicolumn{3}{c}{[.40,.60]}&& \multicolumn{3}{c}{[.80,.95]} \\
\cmidrule(lr){3-5} \cmidrule(lr){7-9} \cmidrule(lr){11-13} \cmidrule(lr){15-17}
&$q$ & {\it sup}\sf{Wald}& {\it exp}\sf{LM}&{\it sup}\sf{LM}&&{\it sup}\sf{Wald}& {\it exp}\sf{LM}&{\it sup}\sf{LM}&&{\it sup}\sf{Wald}& {\it exp}\sf{LM}&{\it sup}\sf{LM}&&{\it sup}\sf{Wald}& {\it exp}\sf{LM}&{\it sup}\sf{LM}\\
 \cmidrule(lr){1-17}
\multirow{ 4}{*}{\rotatebox[origin=c]{90}{\sf SHA}}	&	9	&	25.150	&	0.650	&	0.800	&	&	8.850	&	2.751	&	1.451	&	&	10.070	&	1.301	&	0.600	&	&	25.665	&	14.007	&	4.302	\\
	&	6	&	84.800	&	8.304	&	14.257	&	&	66.310	&	24.762	&	30.665	&	&	67.340	&	9.205	&	10.805	&	&	54.160	&	48.174	&	34.567	\\
	&	3	&	23.800	&	2.401	&	10.205	&	&	69.755	&	17.359	&	30.715	&	&	32.325	&	4.502	&	6.303	&	&	17.880	&	18.409	&	9.955	\\
	&	1	&	15.000	&	0.400	&	1.901	&	&	40.705	&	5.153	&	4.352	&	&	5.470	&	0.900	&	1.201	&	&	14.680	&	11.956	&	1.751	\\
 \cmidrule(lr){3-17}
 \multirow{ 4}{*}{\rotatebox[origin=c]{90}{\sf HAN}}	&	9	&	12.350	&	0.000	&	0.000	&	&	5.445	&	0.000	&	0.000	&	&	48.680	&	1.351	&	1.151	&	&	36.140	&	4.202	&	4.752	\\
	&	6	&	6.100	&	0.000	&	0.000	&	&	2.440	&	0.000	&	0.000	&	&	9.475	&	0.650	&	0.450	&	&	38.460	&	5.103	&	2.901	\\
	&	3	&	0.000	&	0.000	&	0.000	&	&	0.105	&	0.000	&	0.000	&	&	1.480	&	0.500	&	0.100	&	&	15.945	&	2.301	&	0.950	\\
	&	1	&	0.150	&	0.000	&	0.000	&	&	0.150	&	0.000	&	0.000	&	&	4.855	&	0.000	&	0.000	&	&	15.205	&	0.250	&	0.050	\\
 \cmidrule(lr){3-17}									
\multirow{ 4}{*}{\rotatebox[origin=c]{90}{\sf KOS}}	&	9	&	27.100	&	9.405	&	10.305	&	&	9.560	&	0.200	&	0.250	&	&	94.870	&	40.770	&	25.713	&	&	78.880	&	85.343	&	93.947	\\
	&	6	&	4.200	&	5.553	&	9.155	&	&	1.325	&	0.100	&	0.800	&	&	85.910	&	26.013	&	18.759	&	&	67.860	&	70.585	&	91.196	\\
	&	3	&	0.350	&	2.951	&	2.551	&	&	0.540	&	0.100	&	0.800	&	&	47.815	&	12.456	&	6.653	&	&	99.625	&	78.339	&	75.088	\\
	&	1	&	3.350	&	1.401	&	2.501	&	&	0.965	&	0.050	&	0.350	&	&	45.500	&	4.952	&	7.454	&	&	82.990	&	39.020	&	51.876	\\
 \cmidrule(lr){3-17}				
 \multirow{ 4}{*}{\rotatebox[origin=c]{90}{\sf NIK}}	&	9	&	23.050	&	0.000	&	0.000	&	&	8.005	&	0.000	&	0.000	&	&	21.185	&	0.000	&	0.000	&	&	91.845	&	9.105	&	5.303	\\
	&	6	&	4.200	&	0.000	&	0.000	&	&	6.755	&	0.000	&	0.000	&	&	6.165	&	0.000	&	0.000	&	&	97.065	&	6.503	&	6.703	\\
	&	3	&	0.200	&	0.000	&	0.000	&	&	0.125	&	0.000	&	0.000	&	&	0.495	&	0.000	&	0.000	&	&	79.005	&	2.051	&	0.500	\\
	&	1	&	0.000	&	0.000	&	0.000	&	&	0.005	&	0.000	&	0.000	&	&	0.560	&	0.000	&	0.000	&	&	66.245	&	0.450	&	0.300	\\
 \cmidrule(lr){3-17}		
\multirow{ 4}{*}{\rotatebox[origin=c]{90}{\sf SNP}}	&	9	&	49.600	&	7.554	&	29.115	&	&	41.165	&	10.605	&	11.056	&	&	58.780	&	30.515	&	35.018	&	&	19.355	&	1.401	&	1.151	\\
	&	6	&	85.900	&	51.576	&	72.636	&	&	64.760	&	13.007	&	6.653	&	&	90.725	&	92.096	&	84.692	&	&	50.935	&	19.460	&	41.521	\\
	&	3	&	23.700	&	54.927	&	33.217	&	&	28.275	&	9.805	&	10.305	&	&	98.700	&	93.397	&	96.298	&	&	86.275	&	45.773	&	74.387	\\
	&	1	&	62.300	&	66.183	&	64.932	&	&	95.250	&	87.794	&	96.648	&	&	59.255	&	61.281	&	60.130	&	&	30.890	&	11.056	&	13.457	\\
 \bottomrule
\end{tabular}
\end{adjustbox}
\caption{$p$-values (percentages) for detecting Granger causality from lagged WTI returns ($z_i$) to different stock index returns ($y_i$).}
\label{TableApp1}
\end{center}
\end{table}

\begin{table}[!ht]
\begin{center}
\begin{adjustbox}{max width=\textwidth}
\begin{tabular}{lcccccccccc}\toprule
$q$ & \multicolumn{2}{c}{\sf{SHA}} &\multicolumn{2}{c}{\sf{HAN}} & \multicolumn{2}{c}{\sf{KOS}} & \multicolumn{3}{c}{\sf{NIK}}  \\
 \cmidrule(r){1-10}
  \\[-17.5pt]
   \multirow{ 4}{*}{9} & \multicolumn{2}{c}{\footnotesize (0.650)} &  \multicolumn{2}{c}{{\footnotesize (0.000)}} & \multicolumn{2}{c}{{\footnotesize (9.405)}} & \multicolumn{3}{c}{{\footnotesize (0.000)}}  \\
  & 1996-01-03 & 2009-03-06  & 1996-01-03 & 2009-02-20 & 1996-01-03 & 2009-03-17 & 1996-01-03& 2008-09-03 & 2009-08-20    \\[-2pt]
 & 2009-03-06 & 2012-10-12 &  2009-02-20 & 2012-10-12 & 2009-03-17 & 2012-10-12 & 2008-09-03& 2009-08-20 & 2012-10-12  \\
  &  \xmark & \cmark &\xmark & \cmark & \xmark & \cmark & \xmark & \cmark & \cmark \\
  \cmidrule(lr){2-3}  \cmidrule(lr){4-5} \cmidrule(lr){6-7} \cmidrule(lr){8-10}  \\[-17.5pt]
   \multirow{ 4}{*}{6} &\multicolumn{2}{c}{{\footnotesize  (8.304)}}  &  \multicolumn{2}{c}{\footnotesize (0.000)} & \multicolumn{2}{c}{\footnotesize (5.553)} & \multicolumn{3}{c}{\footnotesize (0.000)}  \\
 & 1996-01-03 & 2009-03-06  & 1996-01-03 & 2009-02-20 & 1996-01-03 & 2009-03-31 & 1996-01-03& 2008-09-26 & 2009-08-20    \\[-2pt]
 & 2009-03-06 & 2012-10-12 &  2009-02-20 & 2012-10-12 & 2009-03-31 & 2012-10-12 & 2008-09-26& 2009-08-20 & 2012-10-12  \\
   &  \xmark & \cmark &\xmark & \cmark & \xmark & \cmark & \xmark & \cmark & \cmark \\
  \cmidrule(r){2-3}  \cmidrule(r){4-5} \cmidrule(r){6-7} \cmidrule(r){8-10}  \\[-17.5pt]
   \multirow{ 4}{*}{3} & \multicolumn{2}{c}{\footnotesize (2.401)} &  \multicolumn{2}{c}{\footnotesize (0.000)} & \multicolumn{2}{c}{\footnotesize (2.951)} & \multicolumn{3}{c}{\footnotesize (0.000)}  \\
 & 1996-01-03 & 2009-02-17  & 1996-01-03 & 2009-02-20 & 1996-01-03 & 2009-03-31 & 1996-01-03& 2008-09-18 & 2009-08-24    \\[-2pt]
 & 2009-02-17 & 2012-10-12 &  2009-02-20 & 2012-10-12 & 2009-03-31 & 2012-10-12 & 2008-09-18& 2009-08-24 & 2012-10-12  \\
  &  \xmark & \cmark &\xmark & \cmark & \xmark & \cmark & \xmark & \cmark & \cmark \\
    \cmidrule(r){2-3}  \cmidrule(r){4-5} \cmidrule(r){6-7} \cmidrule(r){8-10}  \\[-17.5pt]
   \multirow{ 4}{*}{1} & \multicolumn{2}{c}{\footnotesize (0.400)} &  \multicolumn{2}{c}{\footnotesize (0.000)} & \multicolumn{2}{c}{\footnotesize (1.401)} & \multicolumn{3}{c}{\footnotesize (0.000)}  \\
 & 1996-01-03 & 2008-10-19  & 1996-01-03 & 2008-10-13 & 1996-01-03 & 2008-10-13 & 1996-01-03& 2008-09-18 & 2009-08-24    \\[-2pt]
 & 2008-12-19 & 2012-10-12 &  2008-10-13 & 2012-10-12 & 2008-10-13 & 2012-10-12 & 2008-09-18& 2009-08-24 & 2012-10-12  \\
  &  \xmark & \cmark &\xmark & \cmark & \xmark & \cmark & \xmark & \cmark & \cmark \\
 \bottomrule
\end{tabular}
\end{adjustbox}
\caption{Identified regimes of Granger causality (checkmark) or no Granger causality (cross) for the interval $\tau \in [0.05,0.95]$ based on Algorithm 2 using an initial significance level of ten percent together with the $p$ values (percentage) of the ${\it exp}{\sf LM}$ test over the complete time series $\{1,\dots,n\}$ for lag length $q \in \{1,3,6,9\}$. }\label{emp:regimes}
\end{center}
\end{table}

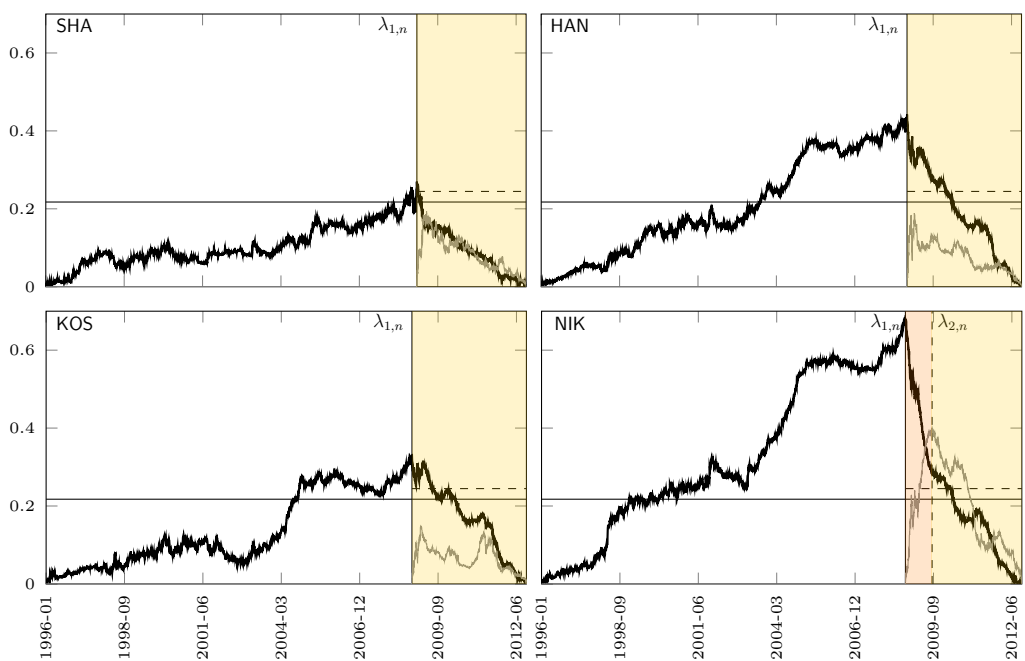
\begin{figure}[!ht]
\centering
\begin{subfigure}[b]{.5\textwidth}
\centering  
  \pgfplotsset{scaled y ticks=false}
\begin{tikzpicture}
        \begin{axis}[
            date coordinates in=x,
            date ZERO=1996-01-03,   
            xmin=1996-01-03,
            xmax=2012-10-12,
            ymin = 0, ymax = .7,	
                yticklabel style={
        /pgf/number format/fixed,
        /pgf/number format/precision=5
},
  width=1.075\textwidth,
    height=.25\textheight,
tick label style={font=\tiny},
xticklabels=\empty,  
        ]
\addplot[solid,thick] table [x=date, y=cusum] {plot/SHA_SEQ.txt};
\draw[solid] (axis cs:2008-12-19,0) -- (axis cs:2008-12-19,.7);
\draw[solid] (axis cs:1996-01-03,0.2174302) -- (axis cs:2012-10-12,0.2174302 );
\draw[dashed] (axis cs:2008-12-19,0.2446425 ) -- (axis cs:2012-10-12,0.2446425);
\addplot[black!45!white,solid] table [x=date,y=cusumB]{plot/SHA_SEQ.txt};
 \node[scale=0.6, align=left]  at (axis cs:2008-02-19, 0.67)  {\(\lambda_{1,n}\)};
           \node[scale=0.6, align=right]  at (axis cs:1997-1-03, .67)  {{\sf SHA}};
  \fill [fill=red!20!yellow, opacity=0.2]
      (axis cs:2008-12-18,0)
      -- (axis cs:2008-12-18,.7)
      -- (axis cs:2012-10-12,.7)
       -- (axis cs:2012-10-12,0)
       -- cycle;
         \end{axis}
   \end{tikzpicture}
\end{subfigure}\hspace*{-.8cm}
\begin{subfigure}[b]{.5\textwidth}
\centering 
  \pgfplotsset{scaled y ticks=false}
\begin{tikzpicture}
        \begin{axis}[
            date coordinates in=x,
            date ZERO=1996-01-03,   
            xmin=1996-01-03,
            xmax=2012-10-12,
            ymin = 0, ymax = .7,	
                yticklabel style={
        /pgf/number format/fixed,
        /pgf/number format/precision=5
},
  width=1.075\textwidth,
    height=.25\textheight,
tick label style={font=\tiny},
yticklabels=\empty,  
xticklabels=\empty,  
        ]
\addplot[solid,thick] table [x=date, y=cusum] {plot/HAN_SEQ.txt};
\draw[solid] (axis cs:2008-10-13,0) -- (axis cs:2008-10-13,.7);
\draw[solid] (axis cs:1996-01-03,0.2174302) -- (axis cs:2012-10-12,0.2174302 );
\draw[dashed] (axis cs:2008-10-13,0.2446425 ) -- (axis cs:2012-10-12,0.2446425);
\addplot[black!45!white,solid] table [x=date,y=cusumB]{plot/HAN_SEQ.txt};
 \node[scale=0.6, align=left]  at (axis cs:2007-12-19, 0.67)  {\(\lambda_{1,n}\)};
           \node[scale=0.6, align=right]  at (axis cs:1997-1-03, .67)  {{\sf HAN}};
  \fill [fill=red!20!yellow, opacity=0.2]
      (axis cs:2008-10-13,0)
      -- (axis cs:2008-10-13,.7)
      -- (axis cs:2012-10-12,.7)
       -- (axis cs:2012-10-12,0)
       -- cycle;
         \end{axis}
   \end{tikzpicture}
\end{subfigure} 
\hspace*{.075cm}
\begin{subfigure}[t]{.5\textwidth}
\centering  
  \pgfplotsset{scaled y ticks=false}
\begin{tikzpicture}
        \begin{axis}[
            date coordinates in=x,
            date ZERO=1996-01-03,   
            xmin=1996-01-03,
            xmax=2012-10-12,
            ymin = 0, ymax = .7,	
                yticklabel style={
        /pgf/number format/fixed,
        /pgf/number format/precision=5
},
  width=1.075\textwidth,
    height=.25\textheight,
tick label style={font=\tiny},
            xticklabel style={
                rotate=90,
                anchor=near xticklabel,
            },
            xticklabel=\year-\month,
        ]
\addplot[solid,thick] table [x=date, y=cusum] {plot/KOS_SEQ.txt};
\draw[solid] (axis cs:2008-10-13,0) -- (axis cs:2008-10-13,.7);
\draw[solid] (axis cs:1996-01-03,0.2174302) -- (axis cs:2012-10-12,0.2174302 );
\draw[dashed] (axis cs:2008-10-13,0.2446425 ) -- (axis cs:2012-10-12,0.2446425);
\addplot[black!45!white,solid] table [x=date,y=cusumB]{plot/KOS_SEQ.txt};
 \node[scale=0.6, align=left]  at (axis cs:2007-12-19, 0.67)  {\(\lambda_{1,n}\)};
           \node[scale=0.6, align=right]  at (axis cs:1997-1-03, .67)  {{\sf KOS}};
  \fill [fill=red!20!yellow, opacity=0.2]
      (axis cs:2008-10-13,0)
      -- (axis cs:2008-10-13,.7)
      -- (axis cs:2012-10-12,.7)
       -- (axis cs:2012-10-12,0)
       -- cycle;
         \end{axis}
   \end{tikzpicture}
\end{subfigure}\hspace*{-.8cm}
\begin{subfigure}[t]{.5\textwidth}
\centering 
  \pgfplotsset{scaled y ticks=false}
\begin{tikzpicture}
        \begin{axis}[
            date coordinates in=x,
            date ZERO=1996-01-03,   
            xmin=1996-01-03,
            xmax=2012-10-12,
            ymin = 0, ymax = .7,	
                yticklabel style={
        /pgf/number format/fixed,
        /pgf/number format/precision=5
},
  width=1.075\textwidth,
    height=.25\textheight,
tick label style={font=\tiny},
yticklabels=\empty,  
            xticklabel style={
                rotate=90,
                anchor=near xticklabel,
            },
            xticklabel=\year-\month,
        ]
\addplot[solid,thick] table [x=date, y=cusum] {plot/NIK_SEQ.txt};
\draw[solid] (axis cs:2008-09-18,0) -- (axis cs:2008-09-18,.7);
\draw[solid] (axis cs:1996-01-03,0.2174302) -- (axis cs:2012-10-12,0.2174302 );

\draw[dashed] (axis cs:2009-08-24,0) -- (axis cs:2009-08-24,.7);
\draw[dashed] (axis cs:2008-09-18,0.2446425) -- (axis cs:2012-10-12,0.2446425);
\addplot[black!45!white,solid] table [x=date,y=cusumB]{plot/NIK_SEQ.txt};
 \node[scale=0.6, align=left]  at (axis cs:2008-01-10, .67)  {\(\lambda_{1,n}\)};
  \node[scale=0.6, align=right]  at (axis cs:2010-05-17, .67)  {\(\lambda_{2,n}\)};
           \node[scale=0.6, align=right]  at (axis cs:1997-1-03, .67)  {{\sf NIK}};
  \fill [fill=red!50!yellow, opacity=0.2]
      (axis cs:2008-09-18,0)
      -- (axis cs:2008-09-18,.7)
      -- (axis cs:2009-08-24,.7)
       -- (axis cs:2009-08-24,0)
       -- cycle;
         \fill [fill=red!20!yellow, opacity=0.2]
      (axis cs:2009-08-18,0)
      -- (axis cs:2009-08-18,.7)
      -- (axis cs:2012-10-12,.7)
       -- (axis cs:2012-10-12,0)
       -- cycle;
         \end{axis}
   \end{tikzpicture}
\end{subfigure}
\vspace*{-.25cm}
             \caption{Highlighted regimes of Granger causality detected using Algorithm 2, with $q = 1$, $B =$ 1,999 on $\tau \in [0.05,0.95]$, and initial significance level of ten percent. The solid black/grey lines are the CUSUM curves calculated over the whole sample $[0,1]$ and the sub-sample $[\lambda_{1,n},1]$, respectively; the solid/dashed horizontal line represents the critical value at a significance level of $1-(1-\alpha)^{1/2}$ and $1-(1-\alpha)^{1/5}$, respectively.}\label{fig2}
\end{figure}

In case of the {\sf SHA}, {\sf HAN}, and {\sf KOS}, we identify two regimes: Before the onset of the financial crisis of 2008 there seems to be no evidence of Granger causality, while Granger causality is detected thereafter. This also applies to the {\sf NIK}, where, however, Algorithm 2 identifies three regimes $0 < \lambda_{1,n} < \lambda_{2,n} < 1$. To rule out the possibility of false rejection, we apply an additional refinement step by computing the CUSUM statistics over $[0,\lambda_{2,n}]$ and $[\lambda_{1,n},1]$, both of which clearly reject with $p$-values given by $0.001$ and $0.003$,  respectively, thereby confirming our findings. Finally, Figure \ref{fig2} succinctly summarizes the preceding discussion graphically by plotting the respective CUSUM plots for all four stock indices where {\it Step 1} rejects in case of $q = 1$.

We corroborate our findings with an analysis about the structural stability of the correlation $\rho(i,j) = \corr[y_i,z_{i-1}]$ between the lagged {\sf WTI} ($z$) and the stock index returns ($y$). An application of the test for constant correlations from \citet{wied:2012}, whose assumptions are typically plausible in the context of (stock) returns, shows that all $p$-values are smaller than $0.05$ in the four cases, where the new procedure detects a break. For {\sf HAN} and {\sf NIK}, they are smaller than $0.001$, for Shanghai, the $p$-value is $0.009$, for {\sf KOS}, the $p$-value is $0.014$.

\section{Conclusion}
We have proposed new tests for Granger causality that are robust against structural breaks and compete very well against existing tests. For future research, it might be interesting to consider nonlinear quantile models such as in \citet{troster:18} instead of pure linear ones.  Moreover, one could consider systems of equations: In our setting, we have one cross-section regression equation and test for restrictions in this equation. In our empirical application on stock returns, this leads to five different tests for each interval of quantile levels. Merging the information to one system of equations might increase the power even further because cross-sectional dependence would be taken into account then.

 \bibliography{bibl}
\appendix
\renewcommand{\theequation}{\Alph{section}.\arabic{equation}}
\setcounter{equation}{0}
\section{Proofs}
\renewcommand{\thelemma}{A.1}
\begin{lemma}\label{lemS0} Recall $\beta_0(\tau) = (0_p',\alpha_0(\tau)')'$ is the parameter value under $H_0$, and suppose Assumptions \ref{ass:A}-\ref{ass:C} are satisfied.
\begin{enumerate}
    \item[\textnormal{($i$)}] For a given $\tau \in {\cal T}$ it holds uniformly in $\lambda \in [0,1]$
    \[
 h(\tau)H_n(\lambda,\tau,\beta_0(\tau)) \Rightarrow {\cal B}_m(\lambda)+ {\sf J}^{-1/2}{\sf H}(\tau) R\left(\lambda  \Delta(\tau)+\delta(\tau) \int_0^\lambda g(v){\sf d}v\right).
 \]    
 \item[\textnormal{($ii$)}] If Assumptions \ref{ass:A}-\ref{ass:C} hold uniformly in $\tau$, then uniformly in $(\tau,\lambda) \in {\cal T} \times [0,1]$
     \[
 H_n(\lambda,\tau,\beta_0(\tau)) \Rightarrow {\cal S}_m(\lambda)+ {\sf J}^{-1/2}{\sf H}(\tau) R\left(\lambda  \Delta(\tau)+\delta(\tau) \int_0^\lambda g(v){\sf d}v\right).
 \]    
\end{enumerate}
\end{lemma}

\textbf{Proof of Lemma \ref{lemS0}.} To begin with, we note that (by \citealp[Lem A.1,A.2]{qu:2008}) the process $S_{n}(\lambda,\tau,\beta_{0}(\tau))$ is stochastically equicontinuous on $[0,1]\times {\cal T}$ equipped with the norm $\rho(\{\lambda_1,\tau_1\},\{\lambda_2,\tau_2\}) = |\lambda_2-\lambda_1|+|\tau_2-\tau_1|$. We now prove first Part ($i$). Fix some $\tau \in {\cal T}$, and let $\beta_{i,n}(\tau) \coloneqq (\gamma_{i,n}(\tau)',\alpha_{0}(\tau)')'$ denote the $m\times 1$ parameter vector under the local alternatives given by Eq.\ \eqref{localt}. Now, under the sequence of local alternatives in Eq.\ \eqref{localt}, one gets
\begin{align}
  S_{n}(\lambda, \tau,\beta_{0}(\tau)&) \nonumber\\
  = \,& \frac{1}{\sqrt n}\sum_{i=1}^{\floor{\lambda n}} x_{i} \psi_\tau(y_i-x_{i}'\beta_{0}(\tau)) \nonumber\\
  = \,& \frac{1}{\sqrt n}\sum_{i=1}^{\floor{\lambda n}} x_{i} \psi_\tau(y_i-x_i'\beta_{i,n}(\tau)) \nonumber\\
  \,& +  \frac{1}{\sqrt n}\sum_{i=1}^{\floor{\lambda n}} x_{i} \big[1\{y_i \leq x_{i}'\beta_{0}(\tau)\}  -1\{y_i\leq x_i'\beta_{i,n}(\tau)\}  \nonumber \\& \hspace*{5.35cm} - F_i(x_{i}'\beta_{0}(\tau))+F_i(x_{i}'\beta_{i,
n}(\tau))\big] \nonumber\\
\,& + \frac1{\sqrt n} \sum_{i=1}^{\floor{\lambda n}} x_{i} (F_i(x_i'\beta_{i,n}(\tau))-F_i(x_{i}'\beta_{0}(\tau))) \eqqcolon I + II + III, \label{eq:decA1}
\end{align}
say. We know that, by Assumption \ref{ass:A}, $x_{i} \psi_\tau(y_i-x_i'\beta_{i,n}(\tau))$ is a martingale difference array under the local alternatives so that by the FCLT  $h(\tau){\sf J}^{-1/2}I \Rightarrow {\cal B}_m(\lambda)$. Moreover, the stochastic equicontinuity of $S_n(\cdot)$ yields  $II = o_p(1)$ (cf. \citealp[Lem A.1/A.2]{qu:2008}). Finally, using a Taylor-series expansion and Eq.\ \eqref{localt}, yields
\begin{align*}
III =  \frac1{\sqrt n}\sum_{i=1}^{\floor{\lambda n}} x_ix_i'f_i(x_i'\beta_{i,n}(\tau))R\gamma_{i,n}(\tau) + o_p(1) \Rightarrow \lambda {\sf H}R\Delta(\tau)+\delta(\tau){\sf H}R\int_0^\lambda g(v){\sf d}v,
\end{align*}
which, as $X_n'X_n/n = {\sf J}+o_p(1)$ (cf. Ass.\ \ref{ass:B}), proves the claim. Part ($ii$) follows analogously noting that ${\sf J}^{-1/2}I  \Rightarrow {\cal S}(\lambda,\tau).$  \hfill$\square$

\textbf{Proof of Proposition \ref{corHtilde}.} Recall that $\alpha_{n}(\tau)$ denotes a solution to the constrained optimization defined in Eq.\ \eqref{eq:beta_tilde} (i.e. assuming $\gamma_{i,n}(\tau) = 0$).  Moreover, recall that $\bar R$ is an $m \times k$ selection matrix such that 
\[
\bar R\alpha_{0}(\tau) = \beta_0(\tau) = \begin{bmatrix}
    0_p \\ \alpha_{0}(\tau)
\end{bmatrix}.
\]
Now, using the same argument that lead to Eq.\ \eqref{eq:decA1}, we get
\begin{align*}
S_n(\lambda,\tau,\tilde\beta_n(\tau))=  S_n(\lambda,&\tau,\beta_0(\tau)) \\
\,&  + \frac{1}{\sqrt{n}}\sum_{i=1}^{\floor{\lambda n}}x_i(F_i(w_i'\alpha_n(\tau)-F_i(w_i'\alpha_{0}(\tau))) + o_p(1)\\
\stackrel{(1)}{=} \,& S_n(\lambda,\tau,\beta_0(\tau)) \\
\,&  + \frac{1}{n}\sum_{i=1}^{\floor{\lambda n}}x_iw_i' f_i(w_i'\alpha_{0}(\tau))\sqrt{n}(\alpha_n(\tau)-\alpha_{0}(\tau)) + o_p(1)\\
\stackrel{(2)}{=} \,& S_n(\lambda,\tau,\beta_0(\tau)) \\
\,&  + \frac{1}{n}\sum_{i=1}^{\floor{\lambda n}}x_ix_i'f_i(x_i'\beta_0(\tau)) (\bar R \sqrt{n}(\alpha_n(\tau)-\alpha_{0}(\tau)))  + o_p(1) \\
\stackrel{(3)}{=} \,& S_n(\lambda,\tau,\beta_0(\tau))   +  \lambda{\sf H}(\tau)\sqrt{n}(\tilde\beta_n(\tau)-\beta_0(\tau)) + o_p(1),
\end{align*}
where $\tilde\beta_n(\tau) = \bar R\alpha_{n}(\tau) = (0_p',\alpha_{n}'(\tau))'$. Equation (1) uses a first order Taylor-series expansion (see \citealp[proof of lem. 1]{qu:2008}), (2) is due to $x_ix_i'\bar R = x_iw_{i}'$, and (3) is due to Ass.\ \ref{ass:C} that also defines the $m \times m$ matrix {\sf H}$(\tau)$. Moreover, recall that ${\sf H}_{\alpha}(\tau)$ and ${\sf J}_{\alpha}$ denote the lower-right $k \times k$ block of {\sf H}$(\tau)$ and {\sf J}, respectively, and partition 
\[
S_n(\lambda,\tau,t) = (S_{n,\gamma}(\lambda,\tau,t_\gamma)',S_{n,\alpha}(\lambda,\tau,t_\alpha)')', \quad t_\gamma \in \mathbb{R}^p,\;t_\alpha \in \mathbb{R}^k
\]
to conform with the partitioning $\beta_0(\tau) = (\gamma_{0}(\tau)',\alpha_{0}(\tau)')'$ for $\gamma_0(\tau) = 0_p$. Then,
\[
\sqrt{n}(\tilde\beta_n(\tau)-\beta_0(\tau)) = -\begin{bmatrix} 0_p \\ {\sf H}_{\alpha}^{-1}S_{n,\alpha}(1,\tau,\alpha_{0}(\tau)) \end{bmatrix} + o_p(1),
\]
see also \citet[Proof of Thm 1]{koemach:99} for the partitioning. Moreover, some algebra reveals
\begin{align*}
  &\begin{bmatrix} 0_p \\ {\sf H}_{\alpha}^{-1}S_{n,\alpha}(1,\tau,\beta_0(\tau)) \end{bmatrix}  \\
 & \quad \quad = \begin{bmatrix} 0_{p \times p} & 0_{p \times k} \\  0_{k \times p} & {\sf H}_{\alpha}^{-1}({\sf J}_{\alpha}^{-1/2})^{-1} \end{bmatrix} \begin{bmatrix} 0_p \\ {\sf J}_{\alpha}^{-1/2}S_{n,\alpha}(1,\tau,\alpha_0(\tau)) \end{bmatrix} \\
 & \quad \quad = \begin{bmatrix} 0_{p \times p} & 0_{p \times k} \\  0_{k \times p} & {\sf H}_{\alpha}^{-1}({\sf J}_{\alpha}^{-1/2})^{-1} \end{bmatrix} {\sf J}^{-1/2}S_n(1,\tau,\beta_0(\tau)) \eqqcolon A(\tau) {\sf J}^{-1/2}S_n(1,\tau,\beta_0(\tau)),
\end{align*}
say, where the $m \times m$ matrix $A(\tau)$ has been implicitly defined. Hence,
\begin{align*}
    H_n(\lambda,\tau,\tilde\beta(\tau)) = \,& {\sf J}^{-1/2}S_n(\lambda,\tau,\beta_0(\tau)) - \lambda {\sf C}(\tau)A(\tau){\sf J}^{-1/2}S_n(1,\tau,\beta_0(\tau)) + o_p(1),   
\end{align*}
where we recall ${\sf C}(\tau) = {\sf J}^{-1/2}{\sf H}(\tau)$. The matrix, $P(\tau) = {\sf C}(\tau)A(\tau)$ is idempotent of rank $k$; cf. Eq\, \eqref{project} so that, by Lemma \ref{lemS0}, the distribution of the restricted quantile estimator follows.

Next, by lemma \ref{lemS0}
\begin{align*}
   h(\tau) R'\tilde H_n(\lambda,\tau,&\tilde\beta(\tau))  \Rightarrow  {\cal B}{\cal B}_p(\lambda)\\
   \,&+  h(\tau)\delta(\tau)R'{\sf C}(\tau)R\left((1-\lambda)\int_0^\lambda g(v){\sf d}v  -\lambda \int_\lambda^1 g(v){\sf d}v\right).
\end{align*}
In particular, if $\lambda = 1$, then
\begin{equation}
\begin{split}
   H_n(1,\tau,\tilde\beta(\tau)) = \,&{\sf J}^{-1/2}S_n(1,\tau,\beta_0(\tau)) \\
    \,& \qquad\qquad - {\sf C}(\tau)A(\tau) {\sf J}^{-1/2}S_n(1,\tau,\beta_0(\tau)) + o_p(1) \\
    = \,& (I- P(\tau)){\sf J}^{-1/2}S_n(1,\tau,\beta_0(\tau)) + o_p(1),
\end{split}
\end{equation}
where we recall that under Ass. \ref{ass:D}, $I_m-P(\tau) = RR'$. \hfill$\square$

 Moreover, if $T(\tau)$ is the inverse of the $m \times m$ matrix of eigenvectors of $I_m - P(\tau)$, then, by the Jordan decomposition (see e.g. \citealp[Ex 8.60]{abad:05}), $I_m-P(\tau) = RR'T(\tau)$ (where $T(\tau) = I_m$ if Ass. \ref{ass:A} holds). Therefore, 
\begin{align*}
  h(\tau)R'H_n(1,\tau,\tilde\beta(\tau)) 
\Rightarrow \,&
    T(\tau){\cal B}_p(1) 
 +
h(\tau)
     R'T(\tau) {\sf C}(\tau)R(\Delta(\tau)-\delta(\tau)\int_0^1g(v){\sf d}v) .  
\end{align*}
Noting that $\mathcal{B}(1)$ and $\mathcal{BB}(\lambda)$ are independent because $\cov[\mathcal{B}(1),\mathcal{BB}(\lambda)] = 0$ by construction, this completes the proof. \hfill$\square$

{\bf Proof of Corollary \ref{corH0}.} Follows immediately from Proposition \ref{corHtilde}.  \hfill$\square$

\textbf{Proof of Corollary \ref{cor1}.} The claim follows from the continuous mapping theorem and Proposition \ref{corHtilde}.  \hfill$\square$

\textbf{Proof of Proposition \ref{prop2}.} The proof follows analogously to that of Proposition \ref{corHtilde} and Corollary \ref{cor1} using Lemma \ref{lemS0} ($ii$) and noting that 
$$\frac{1}{\sqrt n}\sum_{i=1}^{\floor{\lambda n}} x_{i} \psi_\tau(y_i-x_i'\beta_{i,n}(\tau)) \Rightarrow {\cal S}_m(\lambda,\tau)$$
uniformly on $\ell^\infty([0,1] \times {\cal T})$ as mentioned in the proof of Lemma \ref{lemS0}.  \hfill$\square$

\textbf{Proof of Corollary \ref{cor4}.} The claim follows from the continuous mapping theorem and Proposition \ref{prop2}.  \hfill$\square$

\textbf{Proof of Proposition \ref{prop:boot}.} Set $\mathcal{X}_n \coloneqq \{(y_i,x_i), 1 \leq  i \leq n\}$ and let $\mathbb{P}^*(\cdot)$ denote the probability measure induced by the empirical distribution of $\mathcal{X}_n$. Begin by observing that $\alpha_{n,b}(\tau)$ solves the constrained quantile regression problem Eq.\ \eqref{eq:beta_tilde} based on the rescaled data 
$$ \{(\pi_{i,b}w_i'\alpha_n(U_{i,b}),\pi_{i,b}w_i), 1\leq i \leq n\},$$
for bootstrap weights $\pi_{i,b} \stackrel{\sf IID}{\sim} \pi =_d {\sf Multinomial}(n,1/n)$.
Based on this observation, we show first
\begin{align}\label{alphabn}
 \sqrt{n}(\alpha_{n,b}(\tau)-\alpha_n(\tau)) =  \sqrt{n}(\alpha_{n}(\tau)-\alpha(\tau)) + o_{\mathbb{P}^*}(1).
\end{align}
 To see that this is true, observe that 
\begin{align*}
    \sqrt{n}(\alpha_{n,b}(\tau)-\alpha_n(\tau)) = \,& - {\sf H}^{-1}_\alpha(\tau)S_{n,b,\alpha}(\tau,\alpha_n(\tau)) + o_{\mathbb{P}^*}(1),
    \end{align*}
where the sub-gradient of the constrained quantile regression problem is given by
    \begin{align*}
S_{n,b,\alpha}(\lambda,\tau,t_\alpha) \coloneqq \,&   \frac1{\sqrt{n}}\sum_{i=1}^{\floor{\lambda n}} w_{i,b}(1\{\hat y_{i,b} \leq w_{i,b}'t_\alpha\}-\tau), \qquad t_\alpha \in \mathbb{R}^p.
\end{align*}
Note that $S_{n,b,\alpha}(\lambda,\tau,\alpha_n(\tau))$ is centred
\begin{align*}
\Ex[S_{n,b,\alpha}(\lambda,\tau,\alpha_n(\tau)) \mid \mathcal{X}_n ] = \,&\frac1{\sqrt{n}}\sum_{i=1}^{\floor{\lambda n}} \Ex[w_{i,b}(1\{\hat y_{i,b} \leq w_{i,b}'t_\alpha\}-\tau) \mid \mathcal{X}_n] \\
= \,& \frac1{\sqrt{n}}\sum_{i=1}^{\floor{\lambda n}} \Ex[\pi]w_i(\Ex[1\{w_i'\alpha_n(U_{i,b})\leq w_i'\alpha_n(\tau)\} \mid \mathcal{X}_n]-\tau).
\end{align*}
Now, using $\Ex[\pi] = 1$ and that $w_i'\alpha_n(\tau)$ is the $\tau$-quantile of $y_i$ conditional on $\mathcal{X}_n$, one gets $\Ex[S_{n,b,\alpha}(\lambda,\tau,\alpha_n(\tau)) \mid \mathcal{X}_n ] =0$. Similarly, it follows 
\[
\cov[S_{n,b,\alpha}(\lambda_1,\tau_1,\alpha_n(\tau_1)),S_{n,b,\alpha}(\lambda_2,\tau_2,\alpha_n(\tau_2)) \mid \mathcal{X}_n] = (\lambda_1 \wedge \lambda_2)(\tau_1 \wedge \tau_2 - \tau_1\tau_2).
\]
We can conclude by the FCLT for {\sf IID} data that, uniformly in $(\lambda,\tau) \in [0,1] \times {\cal T}$,
\[
S_{n,b,\alpha}(\lambda,\tau,\alpha_n(\tau)) = - {\sf H}_\alpha(\tau){\cal S}_p(\lambda,\tau) + o_{\mathbb{P}^*}(1).
\]
This proves Eq.\ \eqref{alphabn}. Next, define
\[
S_{n,b}(\lambda,\tau,\tilde\beta_{n,b}(\tau)) = \frac1{\sqrt{n}}\sum_{i=1}^{\floor{\lambda n}}x_{i,b}\psi_\tau(\widehat y_{i,b}-x_{i,b}'\tilde\beta_{n,b}(\tau)),
\]
where $\tilde\beta_{n,b}(\tau) = (0_p',\alpha_{n,b}(\tau)')'$.   We can then follow the proof of Proposition \ref{corHtilde}, using the above argument in conjunction with the stochastic equicontinuity of the two-parameter process Eq. \eqref{eq:Hn} in $(\lambda, \tau) \in [0,1] \times {\cal T}$, and obtain
\[
S_{n,b}(\lambda,\tau,\tilde\beta_{n,b}(\tau)) = S_{n,b}(\lambda,\tau,\tilde\beta_{n}(\tau)) + \lambda {\sf H}(\tau)\sqrt{n}(\tilde\beta_{n,b}(\tau)-\tilde\beta_n(\tau))     + o_{\mathbb{P}^{*}}(1).
\]
By Eq.\ \eqref{alphabn}, $\sqrt{n}(\tilde\beta_{n,b}(\tau)-\tilde\beta_n(\tau)) = \sqrt{n}(\tilde\beta_{n}(\tau)-\tilde\beta(\tau)) + o_{\mathbb{P}*}(1)$ and, using similar arguments, $S_{n,b}(\lambda,\tau,\tilde\beta_{n}(\tau)) = S_{n}(\lambda,\tau,\tilde\beta(\tau)) + o_{\mathbb{P}*}(1)$. In view of Proposition \ref{prop2} and Corollary \ref{cor4}, this proves Part ($i$). For Part ($ii$), note that under fixed alternatives $\hat c(\alpha) = O_p(1)$ because the way the bootstrap sample is generated enforces the null hypothesis. Therefore, following the same argument used in the proof of Theorem 3 ($ii$) in \cite{rw:13}, Part ($ii$) is proven. \hfill$\square$

\section{Additional Monte Carlo Results}

We consider the following quantile autoregressive distributed lag model
\[
y_i = \gamma_{i,n}'z_i + \alpha'w_i + u_i, \quad u_i \sim \mathcal{N}(0,1),
\]
where the potentially Granger-causal regressor is given by the $p\times 1$ vector $z_i \coloneqq (z_{1,i},z_{2,i})'$ (i.e. $p =2$), with $z_{1,i} = (1/3)z_{1,i-1}+v_i$, $v_i \sim \mathcal{N}(0,1)$, and $z_{2,i} \sim \chi^2(4)$, while $w_i$ is $k \times 1$ and given by $w_i \coloneqq (1,y_{i-1},y_{i-2},i/n,(i/n)^2,w_{1,i})'$ (i.e. $k =5$ so that $m = 7$) for $w_{1,i} \sim \chi^2(3)$ and $\alpha = (0,1/3,1/4,1/2,1/2,1/2)'$. This design is reminiscent of models considered elsewhere in empirical work; see, e.g. \cite{Chuang2009a} or \cite{gebka:13}.

We conduct a size study ($\gamma_{i,n} = (0,0)')$ and a (local) power study ($\sqrt{n}\gamma_{i,n}  = (1,1)'$ for $i < \floor{n/2}$ and zero otherwise) for our new {\it exp}/{\it sup}{\sf LM} tests (i.e. $\gamma = 0_2$), setting $n \in \{150, 300,$ 1,000, 2,000$\}$ and considering the quantile range $[0.05,0.95]$. We use both the asymptotic critical values (valid only under Assumption \ref{ass:D}) as well as our bootstrap procedure from Algorithm 1. As in our simulation study of the main text, we use 2,000 Monte Carlo iterations and $B = 499$ bootstrap replications. The results collected by Table \ref{tab:adMC} confirm the good finite sample properties in terms of size as well as power, with, in case of power, a superior performance of the exponentially weighted statistic. 

\begin{table}[h]
\begin{center} 
\begin{adjustbox}{max width=\textwidth}
    \begin{tabular}{rccccccc}
    \toprule 
  &&&  \multicolumn{2}{c}{\sf size} && \multicolumn{2}{c}{\sf power}\\
  \cmidrule(lr){4-5} \cmidrule(lr){7-8}
   $n$ & & & {\it sup}{\sf LM} &{\it exp}{\sf LM}&&  {\it sup}{\sf LM} &{\it exp}{\sf LM} \\
  \cmidrule(lr){1-8} 
   \multirow{2}{*}{150} & {\it  asy}  & & 4.50 & 4.30 &&  17.1& 20.3 \\
   & {\it boot}  & & 4.30 & 4.50 &&  16.0 & 22.0 \\
     \cmidrule(lr){2-8} 
        \multirow{2}{*}{300} & {\it  asy}  & & 5.15 & 4.55 &&  18.1 & 21.3 \\
   & {\it boot}  & & 4.91 & 5.01 &&  17.5& 20.1 \\
        \cmidrule(lr){2-8} 
        \multirow{2}{*}{1,000} & {\it  asy}  & & 4.75 & 4.85&&  20.8& 25.6 \\
   & {\it boot}  & & 4.37 & 4.40 &&  19.0 & 29.1 \\
           \cmidrule(lr){2-8} 
        \multirow{2}{*}{2,000} & {\it  asy}  & & 4.60 & 4.95&&  20.4& 26.5 \\
   & {\it boot}  & & 4.50 & 4.60 &&  18.5& 24.0 \\
    \bottomrule
    \end{tabular}
    \end{adjustbox}
     \end{center}
        \caption{Monte Carlo rejection rates (percentages) where `{\sf size}' and `{\sf power}' refer to rejection rates at a five percent significance level under the null and local alternatives, respectively. The labels `{\it asy}' and `{\it boot}' refer, respectively, to the use of critical values obtained from the asymptotic approximation and the bootstrap procedure from Algorithm 1.} \label{tab:adMC}
\end{table}

\end{document}